\documentclass[useAMS,usenatbib]{mnras}
\usepackage{graphicx}
\usepackage{color}
\usepackage{amssymb}
\usepackage{epsfig}
\usepackage{soul}
\usepackage{comment}
\newcommand{\lambdabar}{{\mkern0.75mu\mathchar '26\mkern -9.75mu\lambda}}

\title[Ordinary mode pulsar profile]
{On the triple pulsar profiles generated by ordinary mode}
\author[V. S. Beskin, A. Yu. Istomin, A. G. Mikhaylenko]
{V. S. Beskin$^{1,2,3}$\thanks{E-mail:
beskin@lpi.ru}, A. Yu. Istomin$^{2}$ and A. G. Mikhaylenko$^{2}$ \\
$^{1}$P.N.Lebedev Physical Institute, Leninsky prosp., 53, Moscow, 119991, Russia\\
$^{2}$Moscow Institute of Physics and Technology, Dolgoprudny,
Moscow region,Institutsky per. 9, 141700, Russia \\
$^{3}$National Research Center ''Kurchatov Institute'', Kurchatov sqr.~1, Moscow, 123182, Russia}

\begin{document}

\date{Accepted, Received}

\pagerange{\pageref{firstpage}--\pageref{lastpage}} \pubyear{2023}

\maketitle

\label{firstpage}

\begin{abstract}
A detailed study of the refraction of an ordinary wave in the magnetosphere of radio pulsars was carried out. For this, a consistent theory of the generation of secondary particles was constructed, which essentially takes into account the dependence of the number density and the energy spectrum of secondary particles on the distance from the magnetic axis. This made it possible to determine with high accuracy the refraction of the ordinary O-mode in the central region of the outflowing plasma, which makes it possible to explain the central peak of three-humped mean radio profiles. As shown by detailed numerical calculations, in most cases it is possible to reproduce quite well the observed mean profiles of radio pulsars.
\end{abstract}


\section{Introduction}
\label{Sect0}

Over the past 50 years, a huge amount of observational data has been accumulated on the mean profiles of the pulsar radio emission~\citep{L&M, WJ2008, hankinsrankin2010}. So far in many studies, the analysis of the mean profiles was carried out within the framework of the Rotation Vector Model (RVM,  so-called hollow cone model,~\citealt{radhakrishnan69, O&S}), in which, in particular, rectilinear wave propagation in the magnetosphere of a neutron star was assumed. In this case, it is also usually supposed that the polarization properties of radiation are determined by the structure of the magnetic field in the generation region. 

It is clear that within the framework of such a simplified model, many properties of the mean profiles cannot be explained. For this reason, the hollow cone model has undergone significant modifications (see, e.g.,~\citealt{Shitov, blaskiewicz91, D08, Rookyard15}). One of them is related to the effects of wave propagation in the magnetosphere of radio pulsars which can affect the formation of mean profiles. These include the effects of refraction~\citep{B&A, BGI88}, leading to a significant change in the radiation pattern, cyclotron absorption, which can reduce the intensity in different phases $\phi$~\citep{mikhailovskii82, KMQ00, ML04},  as well as the effects of limiting polarization~\citep{C&R, lim1, lim3, wanglaihan2010, andrianovbeskin2010, beskinphilippov2012, WWH14, HBP}.

\begin{table}
\caption{Pulsars with triple mean profiles taken from~\citet{RankinIV}.
Position angle changes $\Delta p.a.$ and inclination angles $\chi$ are taken from~\citet{L&M}. Five triple pulsars are definitely associated with O-mode only.}
\begin{tabular}{ccccccc}
\hline
PSR & $P$ (s) & ${\dot P}_{-15}$ & $\Delta p.a. (^{\circ})$ & $\chi(^{\circ})$ &  Q & mode \\
\hline
B0329$+$54   &  0.71  &  2.05  & 180 &30&  1.0  &  O \\
B1055$-$52   &  0.20  &  5.84  &  90 &$-$&  0.2  &  X  \\
B1237$+$25   &  1.38  &  0.96  & 180 &48&  2.9  &  O,X  \\
B1508$+$55   &  0.74  &  5.00  & 180 &80&  0.8  &  O,X  \\
B1600$-$49   &  0.33  &  1.02  & 180 &$-$&  0.6  &  O,X?\\
B1700$-$32   &  1.21  &  0.66  & 180 &47&  2.9  &  O \\
B1804$-$08   &  0.16  &  0.03  & 180 &47&  1.1  &  O  \\
B1821$+$05   &  0.75  &  0.23  &  90 &28&  2.6  &  O,X?  \\
B2003$-$08   &  0.58  &  0.05  &  90 &13&  3.6  &  O,X? \\
B2045$-$16   &  1.96  &  10.96 & 180 &37&  1.6  &  O \\
B2111$+$46   &  1.01  &  0.71  & 180 &9&  2.3  &  O  \\
B2319$+$60   &  2.26  &  7.04  &  90 &19&  2.2  &  X?  \\
\hline
\end{tabular}
\label{tab0}
\end{table}

One of these unsolved problems is the question of the so-called triple pulsars, i.e. pulsars with a three-humped mean profiles. For some of them, shown in Table 1, 
linear and circular polarization unambiguously indicates that the mean profile is formed by only one orthogonal mode. This follows from the fact that for these pulsars the position angle $p.a.$ of the linear polarization is well defined, and  they have different signs of circular polarization V and the derivative ${\rm d}p.a./{\rm d}\phi$~\citep{andrianovbeskin2010, beskinphilippov2012}. For the sake of completeness, the table also includes pulsars for which the formation of a mean pulse by one single mode is questionable. It is clear that such a triple-peak behavior is difficult to explain in terms of the standard hollow cone model.

Here, however, it is important for us to emphasize the following. Among the pulsars collected in the table, there are five objects in which the average profile is formed by the ordinary O-mode only. Wherein, as we can see, for all these pulsars the position angle within the mean profile changes by an amount of the order of $180^{\circ}$. Therefore, it can be assumed that the formation of the mean profile in these pulsars is associated with the refraction of the O-mode for central passage of the line of sight through the directivity pattern.

Another fact speaks in favor of this interpretation. All of these five pulsars have parameter $Q = 2 \, P^{1.1}{\dot P}_{-15}^{-0.4}$ greater than unity. Here $P$ is pulsar period in seconds, and ${\dot P}_{-15} = 10^{15}{\dot P}$ is its derivative. As was shown by~\citet{BGI93}, this implies that all these pulsars are located near so-called ''death line'' on the $P$--${\dot P}$ diagram. Hence, for these pulsars, in the center of the directivity pattern one should expect a significant decrease in the radiation intensity associated with a decrease in the number density of the emitting plasma~\citep{RS}. And this, in turn, can significantly affect the refraction of the ordinary mode.

In this paper, we discuss in detail the influence of ordinary wave refraction on the formation of the mean profile of the pulsar radio emission. At present, the theory of refraction of the ordinary O-mode in the magnetosphere of radio pulsars has been developed in sufficient detail. Back in 80-ties,~\citet{B&A} showed that in very strong neutron star magnetic field $B \sim 10^{12}$ G the refractive index can be written as
\begin{equation}
n_2 \approx 1 + \frac{\theta_{\rm b}^2}{4}
- \left(<\frac{\omega_{\rm pe}^2}{\gamma^{3}\omega^2}>
+ \frac{\theta_{\rm b}^4}{16}\right)^{1/2}.
\label{1}
\end{equation}
Here $\omega_{\rm pe} = (4 \pi e^2 n_{\rm e}/m_{\rm e})^{1/2}$ is the electron plasma frequency,  $\theta_{\rm b}$ is the angle between wave vector $\bmath{k}$ and external magnetic field $\bmath{B}$, and $\gamma$ is the Lorentz-factor of outflowing plasma. Recently,~\citet{M&B21} have shown that, for superluminal O-mode, the averaging $<\dots>$ over particle energies remains valid for a fairly wide particle energy distribution (when it is necessary to take care putting $v = c$ in the expression $\omega - \bmath{kv}$).

It is clear that the refraction of an ordinary wave substantially depends on the transverse profile of the number density $n_{\rm e}$. In what follows, we write it down in the form
\begin{equation}
n_{\rm e} = \lambda \, g(r_{\perp}, \varphi_{m})\, n_{\rm GJ},
\label{2}
\end{equation}
where $r_{\perp}$ and $\varphi_{m}$ are magnetic polar coordinates at the star surface. Here 
\begin{equation}
n_{\rm GJ} = \frac{|\bmath{\Omega B}|}{2 \pi  c e}
\label{3}
\end{equation}
is the~\citet{GJ} number density giving diminishing $n_{\rm GJ} \propto r^{-3}$ via decreasing of the magnetic field $B$, $\lambda =$ const is the multiplicity, and the profile $g(r_{\perp}, \varphi_{m})$ determines the transverse distribution of the number density. Wherein, we assume that
\begin{equation}
\int g(r_{\perp}, \varphi_{m}) r_{\perp}{\rm d}r_{\perp}{\rm d}\varphi_{m} = \pi R_{0}^2.
 \label{4}
\end{equation}
On the other hand, for the process of generation of secondary particles produced by one primary particle, we will also determine their number by the quantity $\lambda$, which in this case we will call the multiplication parameter.

It must be said that both in the original~\citep{B&A} and in many subsequent~\citep{BGI88, HB14} works devoted to refraction of the ordinary O-mode, the simplest case $g(r_{\perp}, \varphi_{m}) =$ const was considered. The first results on the influence of a strong dependence $g(r_{\perp})$ on $r_{\perp}$  (as was just predicted by the hollow cone model) were obtained by~\citet{P&L1, P&L2}. In particular, it was shown that for the central passage of the line of sight through the directivity pattern, a third central hump may appear in the mean profile of the radio emission. However, in these works as well as in almost all subsequent works~\citep{wanglaihan2010, andrianovbeskin2010, beskinphilippov2012, HBP, GPB20}, profile $g(r_{\perp}, \varphi_{m})$ was modelled very roughly.

Thus, one of the main tasks of this work is to determine accurately the profile $g(r_{\perp}, \varphi_{\rm m})$ giving us the distribution of the number density $n_{\rm e}$ over the polar cap. Despite the fact that the question of particle production has been actively discussed since the early 80s~\citep{dauharding82, GI85, Hib&A, AE2002, Denis, ML2010, Tim2010,  TimArons2013, TimHar2015}, the dependence on the distance to the axis has not yet been determined. For this reason, model profiles have been used so far, which have not relied on any detailed calculations~\citep{P&L1, P&L2, beskinphilippov2012, HBP}. In addition, as one can see from Eqn. (\ref{1}), to determine the value \mbox{$<\omega_{\rm pe}^2/\gamma^3>$} we also need to know the polar cap distribution of the energy of secondary particles. Sect.~\ref{Sect3} will be devoted to this issue. Here we show how one can evaluate the dependence $<\omega_{\rm pe}^2/\gamma^3>$ on $r_{\perp}$ for small $r_{\perp}$ from a simple consideration.

Indeed, it is natural to relate number density of secondary particles to the number of $\gamma$-quanta that are capable of producing a secondary pair above the polar cap. Assuming that the number density of primary particles $n_{\rm prim}$ accelerated near the neutron star surface does not depend on the position on the polar cap, one can write down
\begin{equation}
\lambda \, g(r_{\perp}) \sim \frac{{\cal E}_{\rm rad}}{{\cal E}_{\rm min}}.
\label{5}
\end{equation}
Here ${\cal E}_{\rm rad}$ is the total energy emitted by a primary particle on the length $L$, and ${\cal E}_{\rm min}$ is the characteristic energy of a $\gamma$-quantum, the free path length $l_{0}$ of which (see below)
\begin{equation}
l_{0} \sim \frac{B_{\rm cr}}{B_{0}} \frac{m_{\rm e}c^2}{{\cal E}_{\rm ph}} R_{\rm c}
\label{7}
\end{equation}
does not exceed the radius of a neutron star $R$ (i.e. the scale at which the decay of the magnetic field begins to strongly affect the rate of pair production). Here  \mbox{$B_{\rm cr} = m_{\rm e}^2c^3/e\hbar \approx 4.4 \times 10^{13}$ G} is the critical magnetic field, ${\cal E}_{\rm ph} = \hbar \, \omega$ is a photon energy, and $R_{\rm c}$ is the curvature radius of the magnetic field line. As a result, we obtain for $l_{0} \sim R$  
\begin{equation}
{\cal E}_{\rm min} \sim \frac{B_{\rm cr}}{B_{0}}
\frac{R_{\rm c}}{R} m_{\rm e}c^2.
\label{8}
\end{equation}

As for the total radiated energy  ${\cal E}_{\rm rad}$, there are two limiting cases. As will be shown below (see Figure~\ref{FigO}), for pulsars with sufficiently long periods $P > 1$ s, the energy of primary particles after acceleration remains practically constant. In this case, one can write down
\begin{equation}
{\cal E}_{\rm rad} \sim \frac{{\rm d}{\cal E}_{\rm prim}}{{\rm d}l} L, 
 \label{6}
\end{equation}
where
\begin{equation} 
\frac{{\rm d}{\cal E}_{\rm prim}}{{\rm d}l} 
\sim \frac{e^2}{R_{\rm c}^2}\left(\frac{{\cal E}_{\rm prim}}{m_{\rm e}c^2}\right)^4. 
 \label{9}
\end{equation}
It gives for $L \sim R$ and ${\cal E}_{\rm prim} \approx$ const
\begin{equation}
\lambda \, g(r_{\perp})  \propto R_{\rm c}^{-3}.
\label{10}
\end{equation}
Using now standard evaluation for the curvature radius of the magnetic field line near magnetic axis $R_{\rm c} \sim R^2/r_{\perp}$, we obtain 
\begin{equation}
\lambda \, g(r_{\perp})  \propto r_{\perp}^3.
\label{11}
\end{equation}
On the other hand, in fast pulsars, the primary particles lose almost all the energy acquired in the acceleration region: ${\cal E}_{\rm rad} \sim e \psi$. For $\psi \approx$ const it gives $\lambda \, g(r_{\perp})  \propto r_{\perp}$.

Finally, if we assume, as is usually done within the framework of the Ruderman-Suthereland model, that the secondary plasma is produced above the vacuum gap in the region of zero longitudinal electric field, then, as is well known, the secondary particles, after emission of all transverse energy due to synchrotron radiation, acquire the energy $\gamma_{\pm}m_{\rm e}c^2$, where $\gamma_{\pm} \sim R_{\rm c}/l_{0}$. Substituting again $l_{0} \sim R$ we obtain 
\begin{equation}
\gamma_{\pm} = k \times 100 \, P^{1/2} \left(\frac{R_{0}}{r_{\perp}}\right),
\label{12}
\end{equation}
were $k \approx 1$.
Here we include into consideration that polar cap radius $R_{0} \approx (\Omega R/c)^{1/2}R$. 
It finally gives for slow pulsars
\begin{equation}
<\frac{\omega^2_{\rm pe}}{\gamma^3}> \, \propto r_{\perp}^6.
\label{13}
\end{equation}
As will be shown in Sect.~\ref{Sect3}, all these asymptotic behaviors are indeed realized with 
a good accuracy.

The paper is organized as follows. In Section 2 we discuss the accelerating potential 
which is necessary to fix the energy of primary particles. In particular, we include into 
consideration the general relativistic correction. Further, in Section 3, we determine 
both the spatial and energy distributions for the first (curvature) and second (synchrotron) 
generations. In general, we follow the approach developed by~\citet{Hib&A}. 
We show that the simple relations (\ref{11}) and (\ref{12}) defined 
above for the dependence of the number density and the characteristic energy of secondary 
particles on the distance from the magnetic axis $r_{\perp}$ are satisfied with good 
accuracy. Finally, in Section 4, the results of calculating the profiles for five O-mode pulsars with 
triple profiles listed in Table~\ref{tab0} are presented. Good agreement of the obtained 
results with observational data is shown.

\section{The energy of primary particles}
\label{Sect1}

\subsection{Accelerating potential}

To begin with, let us discuss the electric potential $\psi$ accelerating primary particles in the polar region.
Here, as a starting point, we use the results of recent numerical simulations~\citep{TimHar2015, PhSC15, PhTS20, Cruz22}. Recall that their main difference from the original models~\citep{RS, Arons1982} is that the process of particle production is essentially non-stationary. However, at the same time, the plasma sometimes completely leaves the polar region. This implies that at these moments a vacuum gap appears above the polar cap, as was predicted by~\citet{RS}. It is not surprising, therefore, that the effective accelerating potential determined by~\citet{TimHar2015} coincides with good accuracy with the Ruderman-Sutherland potential. Therefore, in what follows we assume that the vacuum gap model is an adequate approximation for describing the production of particles in the polar regions of a neutron star.

Thus, below we define the potential drop as
\begin{equation}
\psi_{\rm RS} \approx 2 \pi \rho_{\rm GJ} H^{2},   
\label{psi}
\end{equation}
where $H$ is the inner gap height, and
\begin{equation}
    \rho_{\rm GJ} = -\frac{\bmath{\Omega B}}{2\pi c}
\label{GJ}
\end{equation}
is~\citet{GJ} charge density. Note straight away that \mbox{$\rho_{\rm GJ} \propto B \cos\theta_{b}$,} where $\theta_{b}$ is the angle between angular velocity $\bmath{\Omega}$ and magnetic field $\bmath{B}$. Finally, according to~\citet{TimHar2015}, RS height $H_{\rm RS}$ can be written down as
\begin{equation}
    H_{\rm RS} = 1.1 \times 10^4 |\cos\theta_b|^{-3/7}R_{{\rm c},7}^{2/7}P^{3/7}B_{12}^{-4/7} {\rm cm}.
\label{HRS}
\end{equation}
Here and in what follows the magnetic field $B_{12}$ at the magnetic pole is expressed in $10^{12}$ G, pulsar period $P$ in seconds, and curvature radius $R_{{\rm c},7}$ in $10^{7}$ cm. In Eqn. (\ref{HRS}), as in~\citep{Novoselov}, we add the dependence of charge density on the angle $\theta_b$ between magnetic field and rotational axis into consideration. It is easy to do if one change $\Omega$  to $\Omega|\cos\theta_b|$. In what follows, it is the dependence of the curvature radius $R_{\rm c}$ on the distance from the magnetic axis that will allow us to obtain the spatial distribution of the secondary plasma. 

Recall, however, that the RS potential was obtained under the assumption that the gap height $H$ is much less than the radius of the polar cap $R_{0}$. At least, it is clear that this potential cannot be greater than the potential drop  $\psi_{\rm v}$ corresponding to full vacuum  within open magnetic field line region. In any case, the expression (\ref{psi}) cannot be valid for $r_{\perp} = 0$, when $R_{\rm c} \rightarrow \infty$, and for $r_{\perp} \rightarrow R_{0}$ when $\psi_{\rm RS} \neq 0$.

To determine the vacuum potential drop $\psi_{\rm v}$ it is necessary to solve Poisson equation $\nabla^2 \psi = 4 \pi \rho_{\rm GJ}$ with boundary conditions (see~\citealt{BL22} for more detail)
\begin{eqnarray}
    \psi(r=R, \theta, \varphi) & = & 0,
\label{gr21} \\
    \psi(r, \theta = \theta_{0}(r), \varphi) & = & 0, 
\label{gr20}
\end{eqnarray}
($r$, $\theta$, and $\varphi $ are polar coordinates), where for small angles $\theta$
\begin{equation}
    \theta_{0}(r) = \left(\frac{rR_{0}^2}{R^3}\right)^{1/2}.
\end{equation}
Here we use dipole geometry $\bmath{B} = (3(\bmath{nm})\bmath{n} - \bmath{m})/r^3$. As a result, Poisson equation looks like
\begin{eqnarray}
\frac{1}{r^2} \frac{\partial}{\partial r} 
\left(r^2\frac{\partial\psi}{\partial r}\right) 
+ \frac{1}{r^2\sin\theta}\frac{\partial}{\partial \theta}
\left( \sin\theta\frac{\partial \psi}{\partial \theta}\right) 
+ \frac{1}{r^2\sin^2\theta}\frac{\partial^2 \psi}{\partial \varphi^2} 
\nonumber \\
= - 2\frac{\Omega B_{0}}{c} \, \frac{R^3}{r^3}\left(\cos\theta\cos\chi + \frac{3}{2}\sin\theta\sin\varphi\sin\chi\right),
\label{Eq2}
\end{eqnarray}
which gives for electric potential $\psi$~\citep{BL22}
\begin{eqnarray}
&&\psi(r_{\perp}, \varphi_{m}, l)  =  \frac{1}{2} \, \frac{\Omega B_{0}R_{0}^2}{c}\cos\chi \times 
\nonumber \\ 
&&\left[1 - \left(\frac{r_{\perp}}{R_{0}}\right)^{2}  - \sum_{i} c_{i}^{(0)} \left(\frac{l}{R}\right)^{-\lambda_{i}^{(0)}R/R_{0}}J_{0}(\lambda_{i}^{(0)} r_{\perp}/R_{0})\right]  
    \nonumber \\
&&    + \frac{3}{8} \, \frac{\Omega B_{0}R_{0}^3}{cR} 
\sin\varphi_{m} \sin\chi \left[\left( \frac{r_{\perp}}{R_{0}} - \frac{r_{\perp}^3}{R_{0}^3}\right)\left(\frac{l}{R}\right)^{1/2} \right.
\nonumber \\
&&  \left. -\sum_{i} c_{i}^{(1)} \left(\frac{l}{R}\right)^{-\lambda_{i}^{(1)}R/R_{0}}J_{1}(\lambda_{i}^{(1)}r_{\perp}/R_{0})\right].
\label{psipsi}
\end{eqnarray}
Here $\chi$ is the inclination angle between magnetic moment $\bmath{m}$ and rotation axis, $l$ is the distance from the star surface, $\lambda_{i}$ are the zeros of Bessel functions $J_{i}(x)$, and the values $c_{i}^{(0)}$ and $c_{i}^{(1)}$ are the expansion coefficients giving $\sum_{i} c_{i}^{(0)}J_{0}(\lambda_{i}^{0}x) = 1 - x^2$ and $\sum_{i} c_{i}^{(1)}J_{1}(\lambda_{i}^{1}x) = x - x^3$. As a result, we have for vacuum potential drop $\psi_{\rm V}$ 
\begin{eqnarray}
&&\psi_{\rm v}(r_{\perp}, \varphi_{m})  =  \frac{1}{2} \, \frac{\Omega B_{0}R_{0}^2}{c}
\left(1 - \frac{r_{\perp}^2}{R_{0}^{2}}\right) \cos\chi  
    \nonumber \\
&&    + \frac{3}{8} \, \frac{\Omega B_{0}R_{0}^3}{cR} 
\left(\frac{r_{\perp}}{R_{0}} - \frac{r_{\perp}^3}{R_{0}^3}\right) \sin\varphi_{m} \sin\chi.
\label{psipsip}
\end{eqnarray}
Thus, in what follows we put for the potential drop $\psi$ as
\begin{equation}
\psi = \rm{min}(\psi_{\rm RS}, \psi_{\rm v}).    
\label{psipsipsi}
\end{equation}

\subsection{General relativistic correction}
\label{Sect1_3}

As is well-known, the effects of general relativity and, in particular, 
the frame-dragging (Lense-Thirring) effect, under certain conditions, 
can play a significant role in the generation of secondary plasma near 
the polar caps of a neutron star~\citep{B90, MTs92, Harding_1998, PhSC15, PhTS20}.
For this reason, below we estimate all possible corrections that can 
affect the production of secondary particles. For simplicity, we restrict 
ourselves to only the first order in the small parameter $r_{\rm g} / R$, 
where $r_{\rm g} = 2 GM / c^2 $ is the black hole radius of corresponding mass.

Starting from time-independent Maxwell equation in the rotation reference frame 
(see~\citealt{ThMc86} for more detail)
\begin{equation}
\nabla \times (\alpha \bmath{E} + \bmath{\beta} \times \bmath{B}  + \bmath{\beta}_{\rm R} \times \bmath{B}) = 0,
\label{GR1}
\end{equation}
where $\alpha$ is the lapse function ($\alpha^2 \approx 1 - r_{\rm g}/R$), $\bmath{\beta}$ is Lense-Thirring vector 
($\beta^{\varphi} = -\omega$) and $\bmath{\beta}_{\rm R} = \bmath{\Omega} \times \bmath{r}/c$, we obtain
\begin{equation} 
\alpha \bmath{E} + \bmath{\beta} \times \bmath{B}  + \bmath{\beta}_{\rm R} \times \bmath{B} = - \nabla\psi.
\label{GR2}
\end{equation}
For $\rho_{\rm e} = 0$ it gives
\begin{equation}
\nabla \left(\frac{\nabla \psi}{\alpha}\right) = 4 \pi \rho_{\rm GJ},
\label{GR3}
\end{equation}
where now the Goldreich-Julian charge density looks like
\begin{equation}
\rho_{\rm GJ} = -\frac{1}{8 \pi^2} \nabla_{k}\left(\frac{\Omega - \omega}{\alpha c}\nabla^{k} \Psi\right). 
\label{GR4}
\end{equation}

As we see, the first relativistic correction \mbox{$(1 - \omega/\Omega)$} appears
in the expression for $\rho_{\rm GJ}$ where the ratio $\omega/\Omega$ depends on 
neutron star moment of inertia $I_{r} \sim MR^2$:
\begin{equation}
\frac{\omega}{\Omega} = \frac{I_{r}r_{\rm g}}{M R^3}.
\label{GR5}
\end{equation}
Thus, this correction just corresponds to small value $r_{\rm g}/R$ under consideration.
It is nice that the characteristic scale of the changes in all the relativistic corrections 
is $R$, while the scale of change in $\psi$ is $R_{0} \ll R$. Therefore, one can consider 
all relativistic corrections as constants, which allowed us to put $r = R$ in (\ref{GR5}). 

The second relativistic correction appears in the expression for magnetic field flux
\begin{equation}
\Psi = 2\pi |\bmath{m}|\frac{\sin^2\theta}{r}\left(1 + \frac{3}{4}\frac{r_{\rm g}}{r}\right).
\label{GR6}
\end{equation}
As for small angles $\theta$ one can put $\sin\theta = r_{\perp}/r$, i.e., to write down
\begin{equation}
x^2 = \frac{\Psi}{2 \pi |\bmath{m}|}
y^3\left(1 + \frac{3}{4}\frac{r_{\rm g}}{y}\right)^{-1},
\label{GR7}
\end{equation}
where here $x = r_{\perp}$ and $y = r$,  
we obtain for curvature radius $R_{\rm c} \approx 1/y''_{xx}$ the following correction
$R_{\rm c, GR} = K_{\rm cur}R_{\rm c}$, where 
\begin{equation}
K_{\rm cur} = \left(1 - \frac{1}{2}\frac{r_{\rm g}}{R}\right).
\label{GR8}
\end{equation}
Next, for polar cap radius $R_{0, GR} = K_{\rm cap}R_{0}$ we have
\begin{equation}
K_{\rm cap} = \left(1 - \frac{3}{8}\frac{r_{\rm g}}{R}\right).
\label{GR9}
\end{equation}
Finally, Eqn. (\ref{GR3}) looks now like
\begin{equation}
\frac{\alpha^2}{r_{\perp}}\frac{\partial}{\partial r_{\perp}}\left(r_{\perp}\frac{\partial \psi}{\partial r_{\perp}}\right) 
+ \frac{\partial^2 \psi}{\partial z^2} = 
- \frac{2 \Omega B_{0}}{c} \left(1 + \frac{3}{4} \frac{r_{\rm g}}{R}\right)\left(1 - \frac{\omega}{\Omega}\right).
\label{GR10}
\end{equation}
As a result, we obtain for the general relativistic correction for symmetric potential \mbox{$\psi_{GR}(r_{\perp}) =  K_{\psi}\psi(r_{\perp})$} at distances \mbox{$h > R_{0}$} over the star surface
\begin{equation}
 K_{\psi} = \left(1 - \frac{\omega}{\Omega}\right)\left(1 - \frac{r_{\rm g}}{R}\right)^{-1}.
\label{GR11}
\end{equation}

Thus, the effects of general relativity do turn out to be significant in the analysis 
of the formation of secondary particles. However, for simplicity, below we do not write 
out the corresponding modifications; they will only be included into consideration 
in the final results.

\subsection{The energy of a primary particle}

\begin{figure}
		\center{\includegraphics[width=0.9\linewidth]{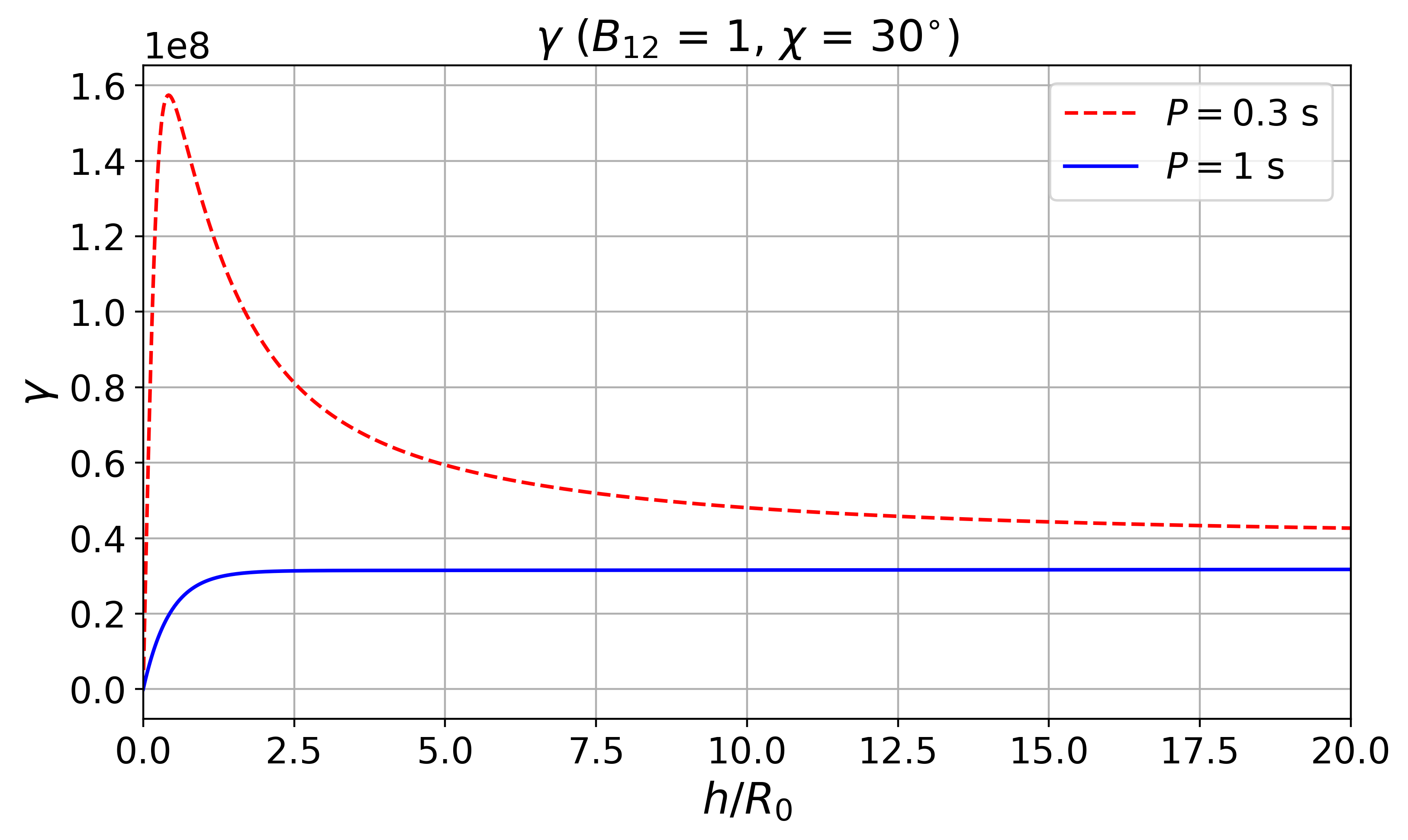}  }
	\caption{Lorentz-factor $\gamma = {\cal E}_{\rm e}/m_{\rm e}c^2$ of a 
	particle accelerated from the surface of a neutron star obtained by 
	solving equation (\ref{motion}) for two values of the pulsar period $P$ 
	for small ($Q < 1$, top) and large ($Q > 1$, bottom) parameter $Q$.}
\label{FigO}	
\end{figure}

Finally, in order to begin discussing the question of the secondary electron-positron 
plasma production, we need to determine the energy of the primary particle 
${\cal E}_{\rm e}$. In general case, the equation of motion looks like 
\begin{equation}
\frac{{\rm d}{\cal E}_{\rm e}}{{\rm d}l} = eE_{\parallel}
-\frac{2}{3} \frac{e^2}{R_{\rm c}^2} \left(\frac{{\cal E}_{\rm e}}{m_{\rm e}c^2}\right)^4.
\label{motion}
\end{equation}
Here $E_{\parallel} = -\partial \psi/\partial l$, and $R_{\rm c}$ again 
is the curvature radius of the magnetic field line. Accordingly, the second 
term is responsible for curvature losses.

Figure~\ref{FigO} shows the dependence of the energy ${\cal E}_{\rm e}$ 
on the distance from the star surface $h$ for the vacuum potential $\psi_{v}$ 
(\ref{psipsi}) for two different pulsar periods $P$ corresponding to small 
($Q < 1$, top) and large ($Q > 1$, bottom) parameter $Q$. As we see, for 
$Q > 1$ the energy losses described by the second term in equation (\ref{motion}) 
becomes negligible, so the condition ${\cal E}_{\rm e}(l) = e \psi(l) - e\psi(l_{0})$ 
($l_{0}$ is the particle starting point) is met with good accuracy.
In this case, the energy of the primary particle reaches a constant value at a height 
$h \approx R_{0}$. For the RS model with $H_{\rm RS} < R_{0}$ this happens at 
even lower altitudes. 

On the other hand, for shorter periods $P$, the particle energy does not reach 
the maximum possible values $e\Delta \psi$, and subsequently decreases with increasing 
the distance from the neutron star surface. However, as shown in Table~\ref{tab0}, 
all the pulsars of interest have the parameter $Q > 1$. On the other hand,
generation of curvature $\gamma$-quanta converting later to secondary pair takes 
place at the distances $h$ up to star radius $R$. For this reason, in our calculations, 
we will neglect both the acceleration region and the dependence of the energy of 
primary particles on the curvature losses.

\section{Generation of secondary pairs}
\label{Sect3}

\subsection{Photon free pass}

As was already noted, in general we follow the approach developed by~\citet{Hib&A}.
Wherein, the main difference is that we obtain the dependence of the energy spectrum 
and multiplicity $\lambda$ of the particle production on the distance $r_{\perp}$
from the magnetic axis.

To determine the free path length of a photon exactly, it is necessary to use 
general expression for the probability $w_{l}$ of photon production at 
a length ${\rm d}l$~\citep{BLP}
\begin{equation}
{\rm d}w_{l} = \frac{3 \sqrt{3}}{16 \sqrt{2}} \, 
\frac{e^3 B\sin\theta_{\rm b}}{\hbar m_{\rm e}c^3}
\exp\left(-\frac{8}{3}\frac{B_{\rm cr}}{B(l)\sin\theta_{\rm b}(l)}
\frac{m_{\rm e}c^2}{{\cal E}_{\rm ph}}\right){\rm d}l.
\label{51}
\end{equation}
Here again $B_{\rm cr} = m_{\rm e}^2c^3/e\hbar \approx 4.4 \times 10^{13}$  G is the critical magnetic field, ${\cal E}_{\rm ph} = \hbar \, \omega$ is a photon energy, and $\theta_{\rm b}$ is the angle between  the wave vector $\bmath{k}$ and the magnetic field $\bmath{B}$. As a result, the free pass length $l_{\gamma}$ is to be determined from the condition
\begin{equation}
\int_{0}^{l_{\gamma}}{\rm d}w_{l} = 1.
\label{52}
\end{equation}

As to photon energy ${\cal E}_{\rm ph}$, in most cases (see, e.g.,~\citealt{TimHar2015, PhSC15, PhTS20}), it was assumed that all the photons emitted by primary particles with the energy ${\cal E}_{\rm e} = \gamma_{\rm e}m_{\rm e}c^2$ are radiated at the characteristic frequency~\citep{LL} 
\begin{equation}
    \omega_{\rm c} = \frac{3}{2} \frac{c}{R_{\rm c}} \gamma_{\rm e}^3.
    \label{41}
\end{equation}
Below we exactly include into consideration the spectrum of the curvature radiation (i.e. the energy radiated in the frequency domain ${\rm d}\omega$ at the distance ${\rm d}l$).
\begin{equation}
    {\rm d}I = \frac{\sqrt{3}}{2 \pi} \frac{e^2}{c R_{\rm c}} 
    \gamma_{\rm e} F(\omega/\omega_{\rm c})  {\rm d} \omega \, {\rm d}l,
    \label{42}
\end{equation}
where
\begin{equation}
    F(\xi) = \xi\int_{\xi}^{\infty}K_{5/3}(x){\rm d}x,
    \label{43}
\end{equation}
$K_{5/3}$ is the Macdonald function, and $\xi = \omega/\omega_{\rm c}$, 
has a rather long tail. As a result, curvature photons will have different 
energies, and therefore different free path lengths $l_{\gamma}$. Moreover, 
as is well known~\citep{sturrock, RS}, relations (\ref{51})--(\ref{52}) give 
with a high accuracy for small enough free pass lengths
$l_{\gamma} \ll R$
\begin{equation}
l_{0} = \frac{8}{3\Lambda} \,R_{\rm c} 
\frac{B_{\rm cr}}{B} \frac{m_{\rm e}c^2}{\hbar \omega}.
    \label{47}
\end{equation}
Here $\Lambda = 15$--$20$ is the logarithmic factor: 
$\Lambda \approx \Lambda_{0} - 3\ln\Lambda_{0}$, where
\begin{equation}
\Lambda_{0} = \ln\left[
\frac{e^2}{\hbar c}\,\frac{\omega_B R_{\rm c}}{c}
\left(\frac{B_{\rm cr}}{B}\right)^2
\left(\frac{m_{\rm e}c^2}{{\cal E}_{\rm ph}}\right)^2\right] \sim 20.
   \label{48}
\end{equation}

\subsection{First (curvature) generation}

To determine the rate of the secondary particle generation, we start 
from one particle moving along magnetic field line intersecting the 
star surface at the distance $r_{0}$ from the axis. Due to (\ref{42}), 
this primary particle produces ${\rm d}N$ photons in the frequency 
domain $ {\rm d}\omega$ at the path ${\rm d}h$
\begin{equation}
 {\rm d}N_{\rm ph}^{(1)} =  \frac{\sqrt{3}}{2 \pi}\frac{e^2}{c R_{\rm c}(h)} 
    \frac{\gamma_{\rm e} F(\omega/\omega_{\rm c})}{\hbar \omega} 
  {\rm d}\omega  \, {\rm d}h.
\label{}
\end{equation}
On the other hand, frequency $\omega$ determines the free path 
$l_{\gamma} = l_{\gamma}(\omega)$ which, in turn,  determines 
the foot point of the magnetic field line $r_{\perp}$ at which the 
secondary pair is created  
\begin{equation}
r_{\perp} = \left(1 - \frac{3}{8} \,\frac{l_{\gamma}^2}{R^2} \right)r_{0}.
\label{xx}
\end{equation}
Note that in a dipole magnetic field, this expansion does not contain 
the corrections $\propto hl_{\gamma}/R^2$ (and, certainly, it does not 
contain the term $\propto h^2/R^2$ as $r_{\perp} = r_{0}$ for $l_{\gamma} = 0$). 

\begin{table}
\caption{Tabulation of the function ${\cal L}(x_{0}, x_{\perp}, h)$ 
(\ref{calL}) for $x_{0} = 0.6$ and for different values $x_{\perp}$.}
\begin{tabular}{ccccccccc}
\hline
$h/R$ & 0.0 & 0.1 & 0.2 & 0.3 & 0.4 & 0.5 & 0.6 & 0.7 \\
\hline
0.599  & 1.0  & 1.4 & 1.9 & 2.6 & 3.4 & 4.3 & 5.4 & 6.7  \\ 
0.59  & 1.5 & 2.0 & 2.7  & 3.4 & 4.3 & 5.4 & 6.6  & 8.1 \\ 
0.58  & 1.9 & 2.5 & 3.3  & 4.1 & 5.1 & 6.3 & 7.7  & 9.3 \\ 
\hline
\end{tabular}
\label{tab2}
\end{table}

As we show below, the leading term in (\ref{xx}) is enough for our consideration. 
On the other hand, in what follows to determine with the required accuracy the 
exponent in the pair creation probability $w_{\rm l}(\theta_{\rm b})$ (\ref{51}) 
we use exact expression for the angle $\theta_{\rm b}$ between magnetic field 
$\bmath{B}$ and the wave vector $\bmath{k}$. In a dipole magnetic field for 
$\gamma$-quanta radiated tangentially at the height $h$ it looks like 
\begin{equation}
\theta_{\rm b} = \frac{3}{4} \, \frac{r_{0}l_{\gamma}}{R^2} f(h),
\label{tb}
\end{equation}
where the correction function can be written down as
\begin{equation}
f(h) = \left(1 + \frac{h}{R}\right)^{1/2}\left(1 + \frac{{\cal L}(h) l_{0}}{R} + \frac{h}{R}\right)^{-1}.
\label{fh}
\end{equation}
Here we introduce by definition another correction function ${\cal L}(h)$ as
\begin{equation}
l_{\gamma}(l_{0}, h)  = {\cal L}(h) \,l_{0},
\label{calL}
\end{equation}
where 
\begin{equation}
l_{0}(\omega) = \frac{32}{9\Lambda} \, \frac{R^2}{r_{0}}
\frac{B_{\rm cr}}{B_{0}} \frac{m_{\rm e}c^2}{\hbar \omega}
\label{}
\end{equation}
is the $\gamma$-quantum free pass in the case $l_{\gamma} \ll R$ with 
the starting point $h = 0$. Coefficient ${\cal L}(h)$ due to the strong 
nonlinearity of the problem for $h \sim l_{0} \sim R$ should be determined
numerically by direct integration (\ref{52}) 
for probability $w_{l}(h)$ corresponding to starting point $h$ at which 
photon free pass is equal to $l_{\gamma}$ (see Table 2). Certainly, 
as one can see, ${\cal L} \rightarrow 1$ for $h \rightarrow 0$ and 
$l_{\gamma} \rightarrow 0$ ($r_{\perp} \rightarrow r_{0}$). Finally, for 
primary particle moving along magnetic field line we have 
\begin{equation}
R_{\rm c} = \frac{4}{3}\, \frac{R^2}{r_{0}} \left(1 + \frac{h}{R}\right)^{1/2}.
\label{Rc}
\end{equation}

Thus, one can write down for the $r_{\perp}$ distribution of the
secondary particles as
\begin{equation}
{\rm d}N_{\pm}^{(1)}  = \frac{\sqrt{3}}{2 \pi} \frac{e^2}{\hbar c} 
    \frac{\gamma_{\rm e} F(\omega/\omega_{\rm c})}{R_{\rm c} \, \omega} 
    \frac{{\rm d}\omega}{\,{\rm d}r_{\perp}}{\rm d}r_{\perp}
 \, {\rm d}h.
\label{xxx}
\end{equation}
To determine the derivative ${\rm d}\omega/{\rm d}r_{\perp}$, one can rewrite 
the relation (\ref{xx}) as
\begin{equation}
\frac{l_{\gamma}(\omega)}{R} = \frac{2\sqrt{2}}{\sqrt{3}} \, \frac{(r_{0}-r_{\perp})^{1/2}}{r_{0}^{1/2}}.
\label{lga}
\end{equation}
It finally gives 
\begin{equation}
\frac{1}{\omega} \frac{{\rm d}\omega}{{\rm d}r_{\perp}} = 
\frac{1}{2(r_{0}-r_{\perp})}\left(1 -  \frac{\omega}{\cal L}\frac{{\rm d}{\cal L}}{{\rm d}\omega}\right)^{-1}.
\label{}
\end{equation}
Within our approximation (\ref{xx}), the value of ${\cal L}$ does not 
depend on $\omega$ (both free path lengths $l_{0}$ and $l_{\gamma}$ 
are mainly determined by the exponent, which both depend on $\omega$ 
as $\omega^{-1}$), and therefore below we do not take into account the
logarithmic derivative $\omega/{\cal L} ({\rm d}{\cal L}/{\rm d}\omega)$.

As a result, we obtain for linear distribution of secondary 
particles ${\rm d}N_{\pm}^{(1)} =  n_{\pm}^{(1)}(x_{\perp}) \, {\rm d}x_{\perp}$ 
created by one primary particle moving along magnetic field line 
with foot point distance from the axis $r_{0}$
\begin{equation}
n_{\pm}^{(1)}(x_{\perp}) = \frac{3\sqrt{3}}{16 \pi} \frac{e^2}{\hbar c}  \frac{R_{0}}{R}  \frac{x_{0}}{(x_{0}-x_{\perp})}
\int_{0}^{H}\frac{{\rm d}h}{R}\gamma_{\rm e}(x_{0},h) F(\xi),
\label{nescd1}
\end{equation}
where now
\begin{equation}
\xi = \frac{64\sqrt{2}}{27\sqrt{3}\Lambda}
\frac{B_{\rm cr}}{B_{0}} \frac{R^3}{\lambdabar R_{0}^2}\frac{\left(1 + {h}/{R} \right)}{\gamma_{\rm e}^3(x_{0},h)}\frac{{\cal L}(x_{0}, x_{\perp}, h)}{x_{0}\sqrt{x_{0}}\sqrt{x_{0}-x_{\perp}}}.
\label{xi1}
\end{equation}
Here we introduce by definition two dimensionless parameters 
\begin{equation}
x_{0} = \frac{r_{0}}{R_{0}}; \qquad x_{\perp} = \frac{r_{\perp}}{R_{0}}.
\label{nesc'}
\end{equation}
Note that expression (\ref{nescd1}) has no singularity at $x_{\perp} = x_0$, since argument $\xi$ (\ref{xi1}) tends to infinity for $x_{\perp} \rightarrow x_0$, so that $F(\xi) \rightarrow 0$.

As for the upper integration limit $H$, it can be set equal to infinity, 
since, as shown in Table~\ref{tab2}, parameter ${\cal L}$ introduced above 
increases rapidly with increasing $h$. Therefore, already at $h \sim R$, 
due to the large value of the argument $\xi$ in (\ref{nescd1}), the 
integrand becomes exponentially small. For this reason in what follows
we do not denote the limits of integration over $h$.

Finally, if the primary particles have 2D spatial distribution 
${\rm d}N_{\rm prim} = n_{\rm prim}(r_{0},\varphi_{m}) r_{0}{\rm d}r_{0}{\rm d}\varphi_{m}$ 
within the polar cap, we obtain for 2D number density of secondary 
pairs ${\rm d}N_{\pm}^{(1)} = n_{\pm}^{(1)}(r_{\perp}, \varphi_{m})r_{\perp}{\rm d}r_{\perp}{\rm d}\varphi_{m}$
\begin{eqnarray}
n_{\pm}^{(1)}  = \frac{\sqrt{3}}{2 \pi} \,\frac{e^2}{\hbar c} \int_{r_{\perp}}^{R_{0}} r_{0}{\rm d}r_{0} \int \, {\rm d}h
\frac{\gamma_{\rm e} F(\omega/\omega_{\rm c})}{R_{\rm c} \omega r_{\perp}} 
    \frac{{\rm d}\omega}{\,\,\,{\rm d}r_{\perp}}n_{\rm prim}.
\label{nesc}
\end{eqnarray}
It gives 
\begin{eqnarray}
&&n_{\pm}^{(1)}(r_{\perp}, \varphi_{m}) = \frac{3\sqrt{3}}{16 \pi} \frac{e^2}{\hbar c}  \frac{R_{0}}{R}  
\nonumber  \\
&&\int_{x_{\perp}}^{1}  \frac{x_{0}^2{\rm d}x_{0}}{x_{\perp}(x_{0}-x_{\perp})}
\int \, \frac{{\rm d}h}{R}\gamma_{\rm e}(x_{0},h) F(\xi)
n_{\rm prim}.
\label{nescd}
\end{eqnarray}

\begin{figure}
		\center{\includegraphics[width=0.9\linewidth]{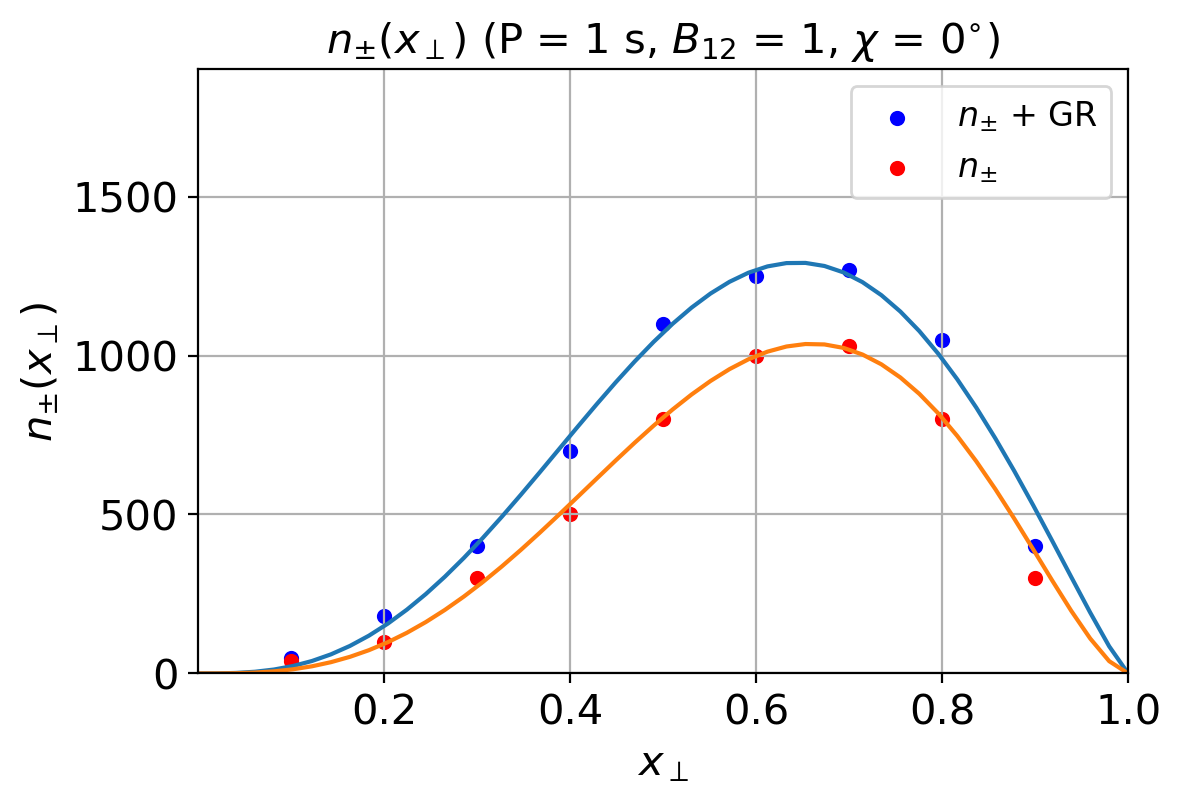}  }
	\caption{2D secondary particle distribution $n_{\pm}(r_{\perp})$ (\ref{nescd}) generated by homogeneous primary particle distribution $n_{\rm prim} = 1$ for ordinary pulsar ($P = 1$ s, $B_{12} = 1$, $\chi = 30^{\circ}$) with (top curve) and without (bottom curve) general relativistic corrections.} 
\label{FigA}	
\end{figure}

On Figure~\ref{FigA} we show 2D secondary particle distribution $n_{\pm}(r_{\perp})$ (\ref{nescd}) generated by homogeneous primary particle distribution $n_{\rm prim} = 1$ for ordinary pulsar ($P = 1$ s, $B_{12} = 1$, $\chi = 30^{\circ}$) with (top curve) and without (bottom curve) general relativistic corrections described in Sect.~\ref{Sect1_3}. Corresponding multiplicities are $\lambda_{\rm GR} = 845$ and $\lambda = 651$. To determine the relativistic corrections, we chose the values $M = 1.4 \, M_{\odot}$ for neutron star mass, $R = 12$ km for that radius, and $I_{r} = 150$ $M_{\odot}$km$^{2}$ for moment of inertia (see~\citealt{Greif20} for more detail). Fitting curves for $x_{\perp} \ll 1$ correspond to asymptotic behavior $n_{\pm} \propto x_{\perp}^3$ (\ref{11}) which, as we can see, perfectly matches the results obtained. 

The approach we have considered above also makes it possible to estimate the energetic spectrum of secondary particles. Let us first consider again the spectrum of secondary particles, which is generated by one primary particle. Then the expression (\ref{xxx}) can be rewritten as
\begin{equation}
{\rm d}N_{\pm}^{(1)}  = \frac{\sqrt{3}}{2 \pi} \frac{e^2}{\hbar c} 
    \frac{\gamma_{\rm e} F(\omega/\omega_{\rm c})}{R_{\rm c} \omega} 
    \frac{{\rm d}\omega}{\,{\rm d}\gamma_{\pm}}{\rm d}\gamma_{\pm}
 \, {\rm d}h.
\label{xxg}
\end{equation}
On the other hand, as one can easily show by passing to a reference frame in which $\gamma$-quantum propagates perpendicular to the external magnetic field, after an almost instantaneous transition to the lower Landau level, the energy of secondary particles can be written as $\gamma_{\pm}m_{\rm e}c^2$ where
\begin{equation}
    \gamma_{\pm} = \frac{1}{\theta_{\rm b}}.
\label{gammapm}
\end{equation}
Hence, expressions (\ref{tb}) and (\ref{gammapm}) can be written down as
\begin{equation}
\gamma_{\pm} = \frac{4}{3} \, \frac{R^2}{r_{0}l_{0}} \, \frac{1}{{\cal L}(h)f(h)},
\label{g1}
\end{equation}
so that
\begin{equation}
\frac{1}{\omega}\frac{{\rm d}\omega}{{\rm d}\gamma_{\pm}} = \frac{1}{\gamma_{\pm}}.
\label{g2}
\end{equation}

As a result, we obtain for the first (curvature) generation energy distribution ${\rm d}N_{\pm}^{(1)}(\gamma_{\pm}) = n_{\pm}^{(1)}(\gamma_{\pm})\,{\rm d}\gamma_{\pm}$ produced by one primary particle with foot point $x_{0}$
\begin{equation}
n_{\pm}^{(1)}(\gamma_{\pm}) = \frac{3\sqrt{3}}{8 \pi} \frac{e^2}{\hbar c}  \frac{R_{0}}{R}    \int \, \frac{{\rm d}h}{R} \, \frac{\gamma_{\rm e}(x_{0},h)}{\gamma_{\pm}}x_{0}
F(\xi),
\label{nescg}
\end{equation}
where
\begin{equation}
\xi = \frac{64}{27\Lambda}
\frac{B_{\rm cr}}{B_{0}} \frac{R^2}{\lambdabar R_{0}}\frac{{\cal L}(h)f(h)(1 + h/R)}{x_{0}\gamma_{\rm e}^3(h)}\gamma_{\pm}.
\label{}
\end{equation}
Accordingly, for the continuous distribution of primary particles $n_{\rm prim}(x_{0}, \varphi)$ we obtain for generation I energy distribution of secondary particles
\begin{equation}
n_{\pm}^{(1)}(\gamma_{\pm}) = \frac{3\sqrt{3}}{8 \pi} \frac{e^2}{\hbar c}  \frac{R_{0}^2}{R^2}    x_{\perp} \gamma_{\pm}^{-1}
\int \, \frac{{\rm d}h}{R}\gamma_{\rm e}(x_{0},h) F(\xi)
n_{\rm prim}.
\label{nescgamma}
\end{equation}

\begin{figure}
		\center{\includegraphics[width=0.9\linewidth]{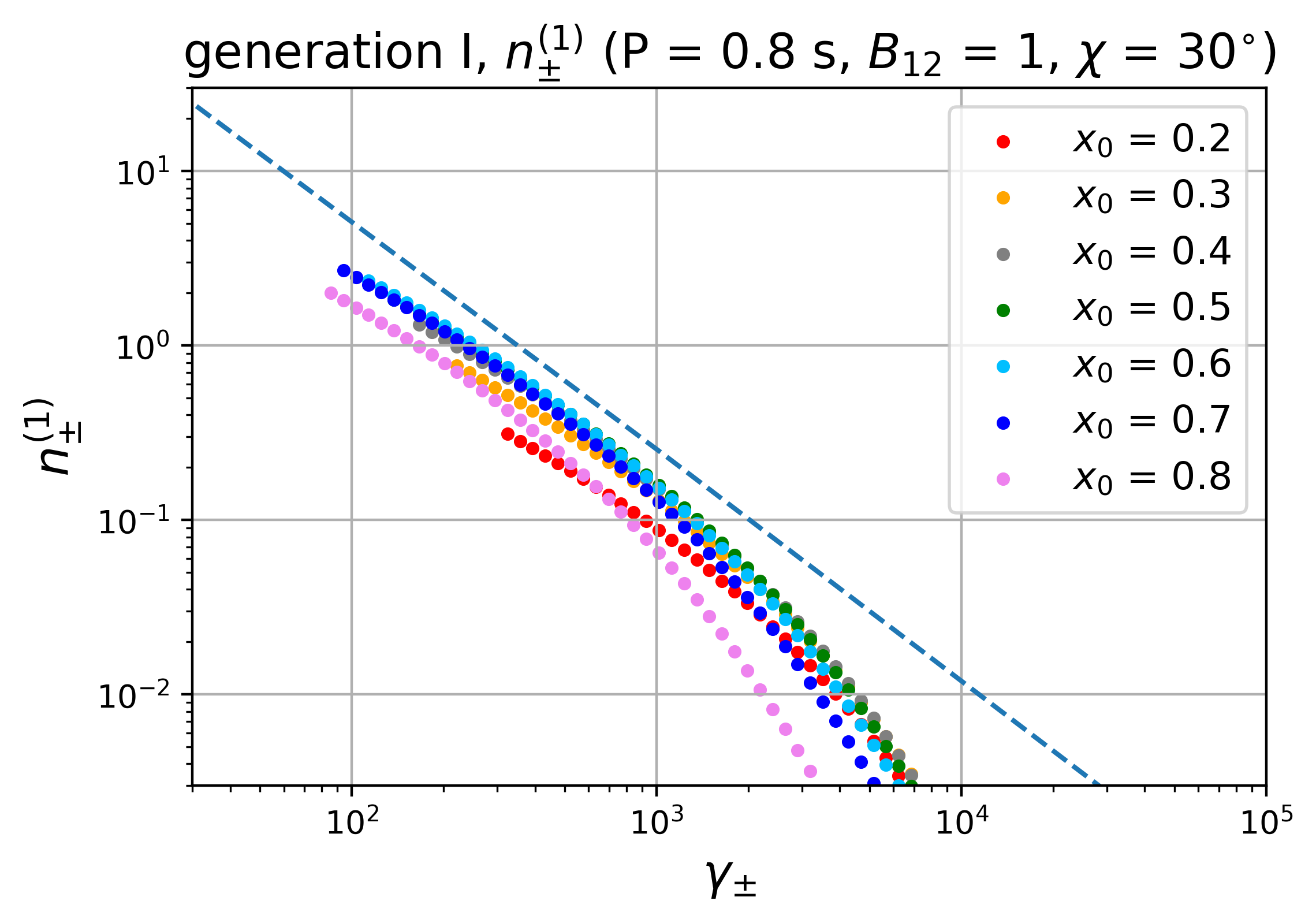}}
	\caption{Energy distribution $n_{\pm}^{(1)}(\gamma_{\pm})$ 
	of generation I secondary particles (\ref{nescgamma})  
	for different distances	$x_{0}$ of the primary particle 
	from magnetic axis for $P = 0.8$ s, magnetic field $B_{12} = 1$, 
	and  $\chi = 30^{\circ}$. Dashed line corresponds to the slope $\gamma_{\pm}^{-4/3}$.}
\label{FigB}	
\end{figure}

\begin{figure}
		\center{\includegraphics[width=0.9\linewidth]{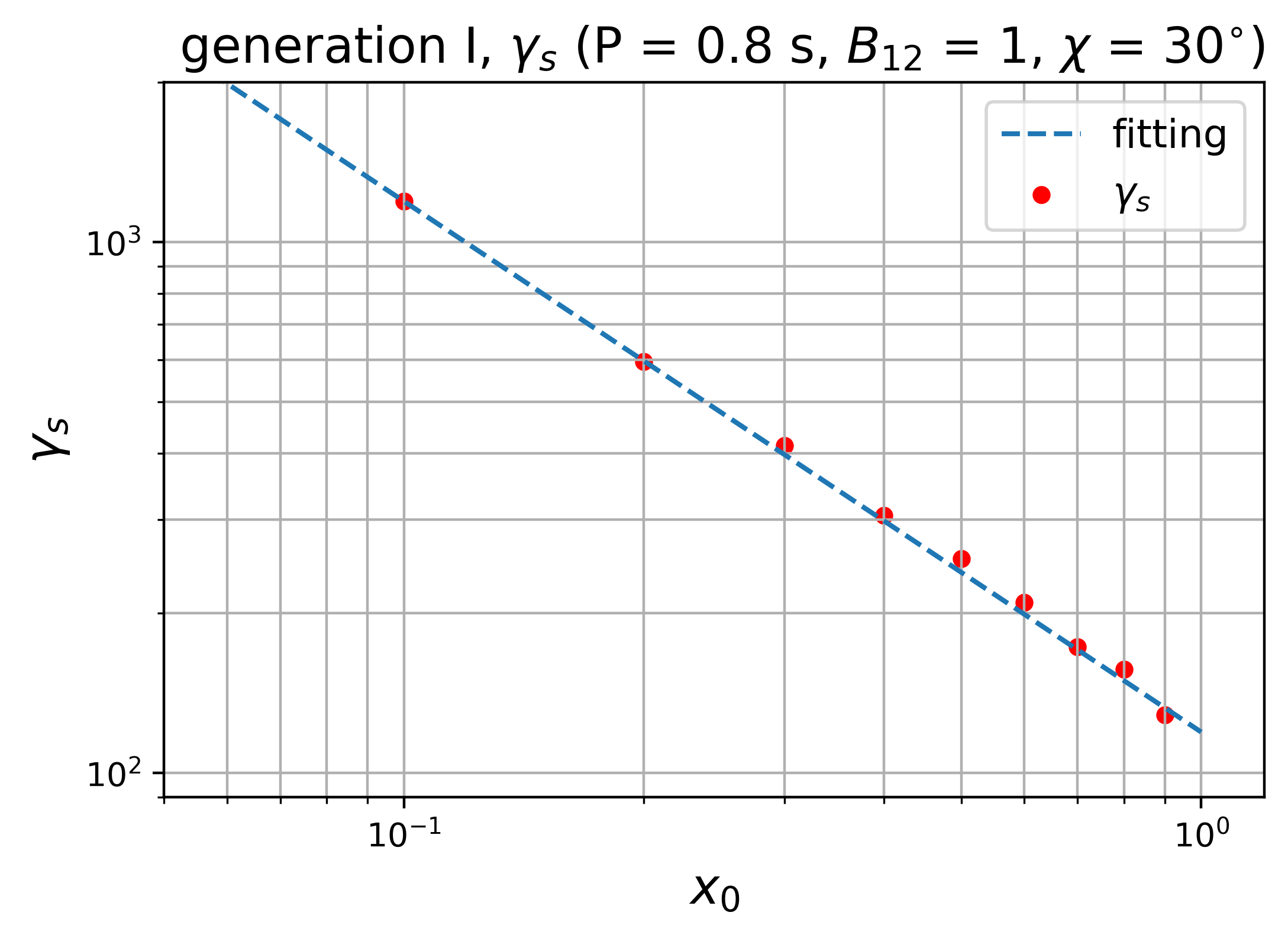}  }
	\caption{Power dependence of the efficient Lorentz-factor \mbox{$\gamma_{\rm s} = <1/\gamma^{3}>^{-1/3}$} 
	on the distance $x_{0}$ of the primary particle from the magnetic axis (generation I). 
	Fitting line corresponds to the slope $\gamma_{\rm s} = 120 \, x_{0}^{-1}$. }
\label{FigC}	
\end{figure}

Figure~\ref{FigB} shows the energy distribution of secondary particles $n_{\pm}^{(1)}(\gamma_{\pm})$ (\ref{nescg})  for different distances $x_{0}$ of the primary particle from magnetic axis for standard values \mbox{$P = 0.8$ s,} magnetic field $B_{12} = 1$, and  $\chi = 30^{\circ}$. It is clearly seen that at low particle energies, the relation $n_{\pm}^{(1)}(\gamma_{\pm}) \propto \gamma_{\pm}^{-4/3}$ is fulfilled (cf.~\citealt{GI85}).  This asymptotic behavior can be easily obtained from relation (\ref{nescgamma}), since $F{\xi} \propto \xi^{1/3}$ for $\xi \ll 1$. We also note a sharp drop in the distribution function at low particle energies, when the magnetosphere becomes transparent for $\gamma$-quanta. As expected, the obtained spectra are quite similar to the spectra obtained by~\citet{Hib&A}.

Further, Figure~\ref{FigC} demonstrates the dependence of the efficient Lorentz-factor \mbox{$\gamma_{\rm s} = <1/\gamma^{3}>^{-1/3}$} on the distance $x_{0}$ of the primary particle from the magnetic axis. Fitting line corresponds to the slope $\gamma_{\rm s} = 120 \, x_{0}^{-1}$ in excellent agreement with power dependence (\ref{12}). Hence, the estimate $<\omega^2_{\rm pe}/\gamma^3 > \, \propto r_{0}^6$ (\ref{13}) remains valid as well. 

\subsection{Second (synchrotron) generation}

As is well known, synchrotron photons emitted by secondary particles play a significant role in the generation of secondary plasma~\citep{dauharding82, GI85, Denis}. The point is that the secondary particles are born with nonzero angles $\theta_{\rm p}$ to the magnetic field. As was shown by~\citet{VBAstrofiz, DH83}, for not so large magnetic field $B_{0} \sim 10^{12}$ G one can use the classical approximation
\begin{equation}
\theta_{\rm p} = \theta_{\rm b},
\label{syn1}
\end{equation}
where $\theta_{\rm b}$ is given by Eqn.  (\ref{tb}). Accordingly, the energies of secondary particles at the moment of their birth $\gamma_{\rm max} m_{\rm e}c^2$ are equal to each other, so that 
\begin{equation}
\gamma_{\rm max} = \frac{1}{2} \, \frac{\hbar \omega}{m_{\rm e}c^2}.
\label{syn2}
\end{equation}

To determine the frequency $\omega^{\prime}$ spectrum of synchrotron photons, it is convenient to start with the the standard form of the single particle emission (see, e.g.,~\citealt{Denis})
\begin{equation}
{\rm d}N_{\rm ph}^{(2)}  = \frac{\sqrt{3}}{2 \pi} \frac{e^2}{\hbar c} \omega_{B} \frac{{\rm d}\omega^{\prime}}{\omega^{\prime}} \int \theta(t)
F(\omega^{\prime}/\omega_{\rm s})
    {\rm d}t,
\label{syn3}
\end{equation}
where now
\begin{equation}
\omega_{\rm s} = \frac{3}{2} \, \omega_{B} \, \theta_{1}(t) \gamma_{1}^2(t),
\label{syn4}
\end{equation}
and subscript 1 corresponds to particles of the first generation. Accordingly, the expression for synchrotron energy losses looks like~\citep{LL}
\begin{equation}
\frac{{\rm d}\gamma_{1}}{{\rm d}t} = - \frac{2}{3} \, \frac{e^2}{m_{\rm e}c^2}\frac{\omega_{B}^2}{c} \, \theta_{1}^2 \gamma_{1}^2.
\label{syn5}
\end{equation}
We remind that the synchrotron radiation time is so short that one can  assume that all synchrotron photons are emitted at the point of production of the secondary pair. 
Using now the ratio between the current values $\gamma_{1}(t)$ and $\theta_{1}(t)$
\begin{equation}
\frac{1}{\gamma_{1}^2} = 1 - \frac{v_{\parallel}^2}{c^2}(1 + \theta_{1}^2),
\label{syn6}
\end{equation}
where $v_{\parallel} = $ const ($1 - v_{\parallel}^2/c^2 = 1/\gamma_{1 \pm}^2$), which gives for $v_{\parallel} \approx c$
\begin{equation}
\theta_{1} \gamma_{1}^2 = \frac{\gamma_{1}}{\gamma_{1 \pm}}(\gamma_{1}^2 - \gamma_{1 \pm}^2)^{1/2},
\label{syn7}
\end{equation}
we have for the number of photons generated by electron-positron pair created by single curvature $\gamma$-quantum having energy $\hbar \omega$
\begin{equation}
{\rm d}N_{\rm ph}^{(2)}  =  2\frac{3\sqrt{3}}{4 \pi} \frac{c}{\lambdabar \omega_{B}}
\frac{{\rm d}\omega^{\prime}}{\omega^{\prime}} \int_{\gamma_{1 \pm}}^{\gamma_{\rm max}}
\frac{\gamma_{1 \pm} F(\omega^{\prime}/\omega_{\rm s})}{\gamma_{1}(\gamma_{1}^2 - \gamma_{1 \pm}^2)^{1/2}}
    {\rm d}\gamma_{1}.
\label{syn8}
\end{equation}
Here the absence of photons with a high frequency $\omega^{\prime}$ for given energy of the curvature $\gamma$-quantum $\hbar \omega$ (which determines the maximum frequency $\omega_{\rm s}$) is associated with a sharp diminishing $F(\omega^{\prime}/\omega_{\rm s})$ at $\omega^{\prime} \gg \omega_{\rm s}$. Accordingly, an additional factor 2 takes into account the fact that a secondary electron-positron pair is involved in the emission of synchrotron photons.

Let us now note the following circumstance, which can significantly simplify our further calculations. Using relation (\ref{syn2}), one can easily find that the ratio of the characteristic frequency of a synchrotron photon $\omega_{\rm s}$ (\ref{syn4}) produced by secondary particle to the frequency of a curvature photon $\omega$ producing this particle
\begin{equation}
\frac{\omega_{\rm s}}{\omega} = \frac{3}{8}\frac{B_{0}}{B_{\rm cr}}
\frac{\hbar \, \omega}{m_{\rm e}c^2}\frac{l_{\gamma}}{R_{\rm c}}
\label{52bis}
\end{equation}
turns out to be exactly $\Lambda^{-1}$. For this reason, the mean free path of synchrotron photons must be $\Lambda$ times greater than the mean free path of curvature photons. Since $\Lambda \gg 1$, we can neglect the difference in the free path length from the direction of the synchrotron radiation of  photons within the radiation cone as they vary from $(\Lambda + 1)\,l_{0}$ to $(\Lambda - 1)\,l_{0}$. In other words, we assume that all synchrotron photons emitted by secondary pair are radiated at the same height $h$ as the curvature $\gamma$-quantum producing this pair. For the same reason, defining $\omega_{\rm B}$  through expression (\ref{syn4}), we can put $B = B_{0}$ (i.e., ignore the correction associated with $h \neq 0$), since synchrotron photons, leading to the production of secondary pairs at $h \sim R$, will be emitted near the surface of a neutron star, and synchrotron photons emitted at $h \sim R$ will not lead to the production of particles.

As a result, we finally obtain for 1D distribution of the second generation particles  ${\rm d}N_{\pm}^{(2)}(x_{\perp})$ produced by single primary particle moving along magnetic field line $x_{0}$
\begin{eqnarray}                                             
&&{\rm d}N_{\pm}^{(2)}(x_{\perp})  =  \frac{27}{32 \, \pi^2} \frac{e^2}{\hbar c} \frac{B_{\rm cr}}{B_{0}}
\frac{R_{0}}{R}
\frac{x_{0} \, {\rm d}x_{\perp}}{(x_{0} - x_{\perp})}  
\label{syn10}   \\
&&
\int \, \frac{{\rm d}h}{R} \gamma_{\rm e}(h, x_{0}) 
\int_{0}^{\infty}\frac{{\rm d}\xi}{\xi} F(\xi)
\int_{1}^{g_{\rm max}}
\frac{F(\xi^{\prime})}{g\sqrt{g^2 - 1}}
    {\rm d}g,
\nonumber
\end{eqnarray}
where we introduce two dimensionless variables $\xi = \omega/\omega_{\rm c}$ and $g = \gamma_{1}/\gamma_{1\pm}$. Accordingly,
$\xi^{\prime} = \xi^{\prime}(g, \xi, x_{\perp}, h)$ can be written down as
\begin{equation}
\xi^{\prime} = \frac{1024\sqrt{2}}{729 \sqrt{3}\Lambda^2} \frac{B_{\rm cr}^3}{B_{0}^3}
\frac{R^2}{\lambdabar R_{0}}\frac{1}{\gamma_{\rm e}^3}\frac{{\cal L}^{2}(x_{0}, x_{\perp}, h)f(h)}{x_{0}\sqrt{x_0}\sqrt{x_{0} - x_{\perp}}}\frac{1}{\xi} \frac{(1 + h/R)^{3/2}}{g\sqrt{g^2 - 1}}.
\label{syn11}
\end{equation}
There is no singularity at $x_{\perp} = x_{0}$ in (\ref{syn10}) as \mbox{$\xi^{\prime}\rightarrow \infty$} for \mbox{$x_{\perp} \rightarrow x_{0}$.}

Further,
\begin{equation}
g_{\rm max} = \frac{4}{3 \Lambda} \frac{B_{\rm cr}}{B_{0}}{\cal L}(x_{0}, x_{\perp}, h)f(h).
\label{syn12}
\end{equation}
Note that the violation of condition $g_{\rm max} > 1$, which occurs at sufficiently 
high magnetic fields, corresponds to the well-known regime of pair production at the 
lower Landau levels~\citep{VBAstrofiz, DH83, Denis}, when the synchrotron radiation of secondary particles actually 
no longer takes place. Therefore, when the condition $g_{\rm max} < 1$ is satisfied, the generation of secondary particles of the second generation does not occur. Finally, as was already noted, 
the upper limit of integration over $h$ is largely determined by the growth of the 
function ${\cal L}(h)$. 

Besides, distribution by Lorentz-factor $\gamma_{\pm}$ looks like
\begin{eqnarray}
&&{\rm d}N_{\pm}^{(2)}(\gamma_{\pm})  =  \frac{27}{16 \, \pi^2} \frac{e^2}{\hbar c} \frac{B_{\rm cr}}{B_{0}}
\frac{R_{0}}{R}
\frac{x_{0} \, {\rm d}\gamma_{\pm}}{\gamma_{\pm}}  
 \label{syn14}   \\
&&
\int \, \frac{{\rm d}h}{R} \gamma_{\rm e}(h, x_{0}) 
\int_{0}^{\infty}\frac{{\rm d}\xi}{\xi} F(\xi)
\int_{1}^{g_{\rm max}}
\frac{F(\xi^{\prime})}{g\sqrt{g^2 - 1}}
    {\rm d}g,
\nonumber
\end{eqnarray}
where again $\xi = \omega/\omega_{\rm c}$, so that
\begin{equation}
\xi^{\prime}(\xi, \gamma_{\pm}) = \frac{1024}{729 \Lambda^2} \frac{B_{\rm cr}^3}{B_{0}^3}
\frac{R^2}{\lambdabar R_{0}}\frac{1}{\gamma_{\rm e}^3}\frac{{\cal L}^{2}(h)f^{2}(h)}{x_{0}}\frac{\gamma_{\pm}}{\xi}
\frac{(1 + h/R)^{3/2}}{g\sqrt{g^2 - 1}}.
\label{syn17} 
\end{equation}

Finally, if the secondary particles have 2D spatial distribution ${\rm d}N^{(1)} = n^{(1)}(r_{0}, \varphi_{m})r_{0}{\rm d}r_{0}{\rm d}\varphi_{m}$ within the polar cap, we obtain for 2D number density of secondary pairs ${\rm d}N^{(2)}_{\pm} = n_{\pm}^{(2)}(r_{\perp}, \varphi_{m})r_{\perp}{\rm d}r_{\perp}{\rm d}\varphi_{m}$
\begin{eqnarray}                                                 
&&n_{\pm}^{(2)}(x_{\perp},\varphi_{m})  =  \frac{27}{32 \, \pi^2} \frac{e^2}{\hbar c} \frac{B_{\rm cr}}{B_{0}}
\frac{R_{0}}{R} \int_{x_{\perp}}^{1}
\frac{n^{(1)} x_{0}^2 \, {\rm d}x_{0}}{x_{\perp}(x_{0} - x_{\perp})}  
 \label{syn18}  \\
&&
\int \, \frac{{\rm d}h}{R} \gamma_{\rm e}(h, x_{0})  
\int_{0}^{\infty}\frac{{\rm d}\xi}{\xi} F(\xi)
\int_{1}^{g_{\rm max}}
\frac{F(\xi^{\prime})}{g\sqrt{g^2 - 1}}
    {\rm d}g,
\nonumber  
\end{eqnarray}
where here $\xi^{\prime}$ is again given by (\ref{syn11}).

\begin{figure}
		\center{\includegraphics[width=0.9\linewidth]{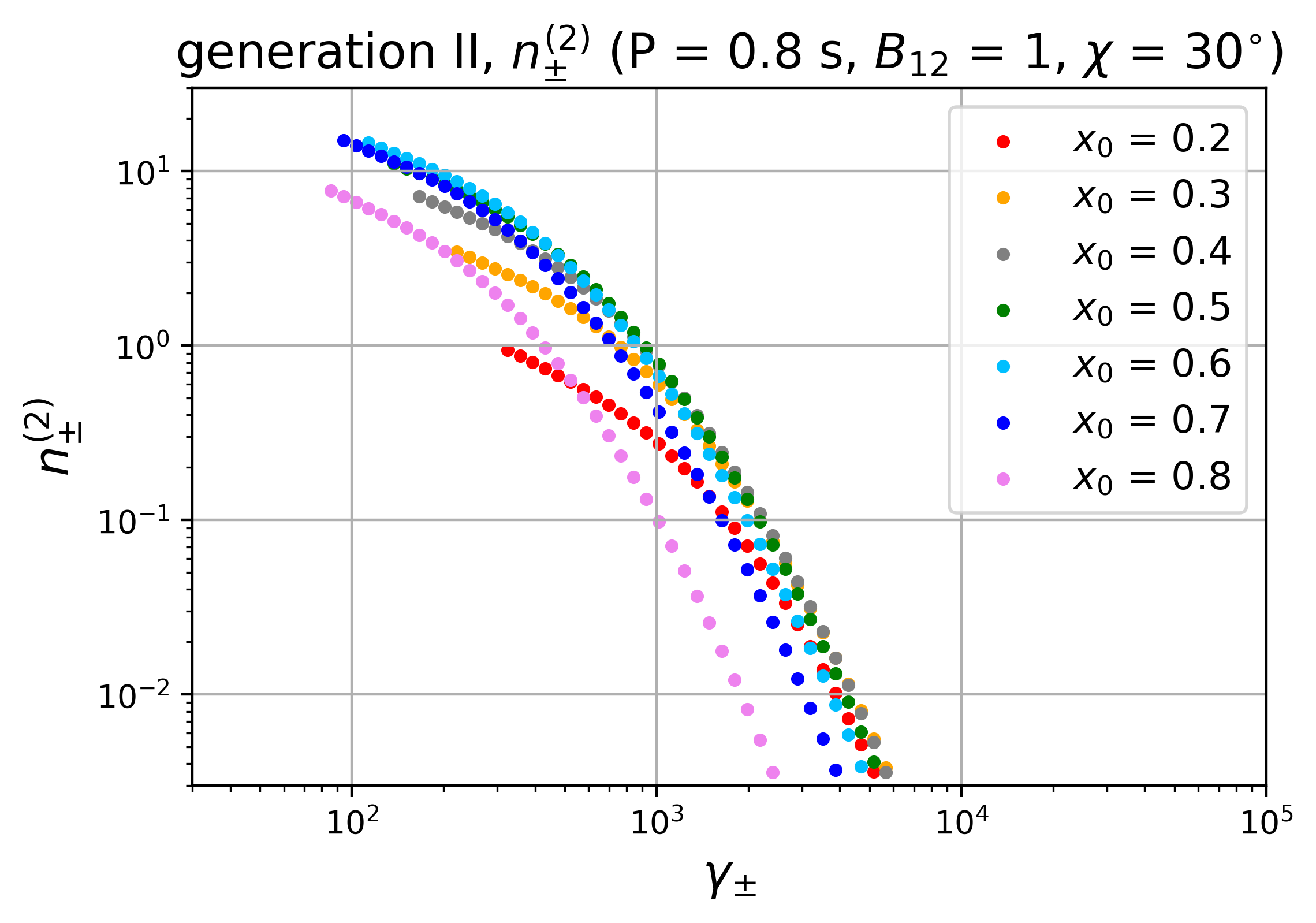}  }
	\caption{Energy distribution $n_{\pm}^{(2)}(\gamma_{\pm})$ 
	of the generation II secondary particles (\ref{nescgamma})  
	for different distances	$x_{0}$ from magnetic axis for 
	$P = 0.8$ s, magnetic field $B_{12} = 1$, 
	and  $\chi = 30^{\circ}$.}
\label{FigD}	
\end{figure}

\begin{figure}
		\center{\includegraphics[width=0.9\linewidth]{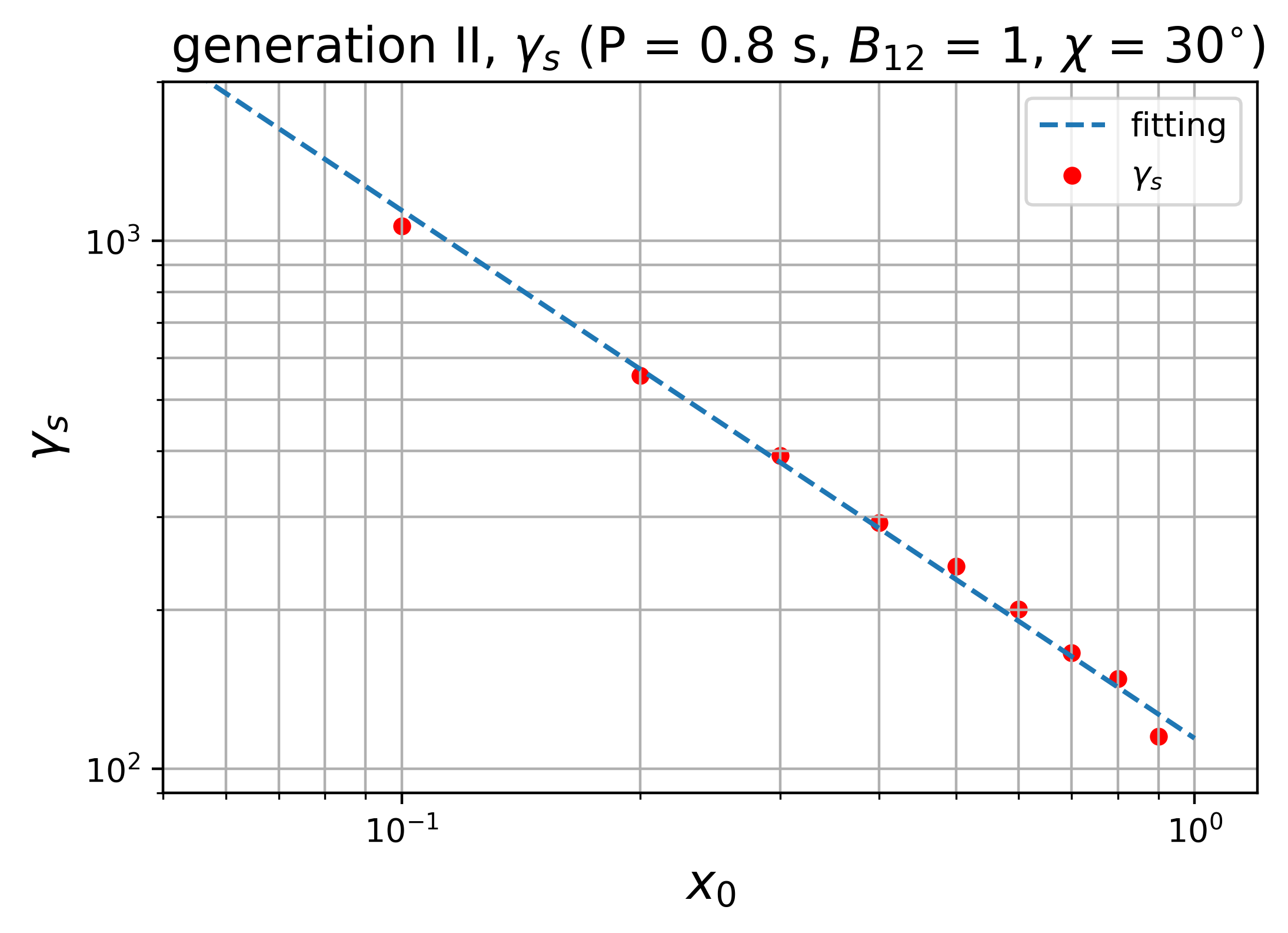}  }
	\caption{Power dependence of the efficient Lorentz-factor \mbox{$\gamma_{\rm s} = 
	<1/\gamma^{3}>^{-1/3}$} on the distance $x_{0}$  (generation II). 
	Fitting line corresponds to the slope \mbox{$\gamma_{\rm s} = 105 \, x_{0}^{-1}$.} }
\label{FigE}	
\end{figure}

Figure~\ref{FigD} shows the energy spectrum of the second (synchrotron) 
generation for the same parameters as in Figure~\ref{FigB}. Accordingly, 
Figure~\ref{FigE} demonstrates the dependence of the efficient Lorentz-factor 
\mbox{$\gamma_{\rm s} = <1/\gamma^{3}>^{-1/3}$} on the distance $x_{0}$ from 
the magnetic axis. As one can see, the power-law dependence 
$n_{\pm}^{(2)} \propto \gamma_{\pm}^{-1}$ is also fulfilled here with
good accuracy. Moreover, this dependence turns out to be quite universal 
and independent of the generation.

\begin{table}
\caption{Multiplication parameters $\lambda_{I}$ and $\lambda_{II}$ for two generations I and II
for different values $x_{0}$ of a primary particle ($B_{12} = 1$).}
\begin{tabular}{ccccccc}
\hline
& $P = 0.8$ s &         &  $P = 1.0$ s  &      &  $P = 1.2$ s &  \\
Gen. &  $\lambda_{I}$   & $\lambda_{II}$ &  $\lambda_{I}$  & $\lambda_{II}$  & $\lambda_{I}$ &  $\lambda_{II}$  \\
\hline
0.1   &   81  &   59 &   2  &    0 & 0 & 0 \\
0.2   &  491  & 1279 &  69  &   50 & 0 & 0 \\
0.3   &  869  & 3598 & 225  &  326 & 4 & 1 \\
0.4   & 1169  & 6047 & 385  &  801 & 14 & 6 \\
0.5   & 1350  & 7801 & 469  & 1126 & 20 & 10 \\
0.6   & 1443  & 8568 & 457  & 1057 & 16 & 8 \\
0.7   & 1356  & 7285 & 320  &  565 &  6 & 2 \\
0.8   &  825  & 2882 &  90  &   82 &  0 & 0 \\
0.9   &   78  &  67  &   0  &    0 &  0 & 0 \\
\hline
total &  803  & 4055 & 209  &  419  & 6 & 3\\
\hline
\end{tabular}
\label{tab3}
\end{table}

\begin{figure}
		\center{\includegraphics[width=0.9\linewidth]{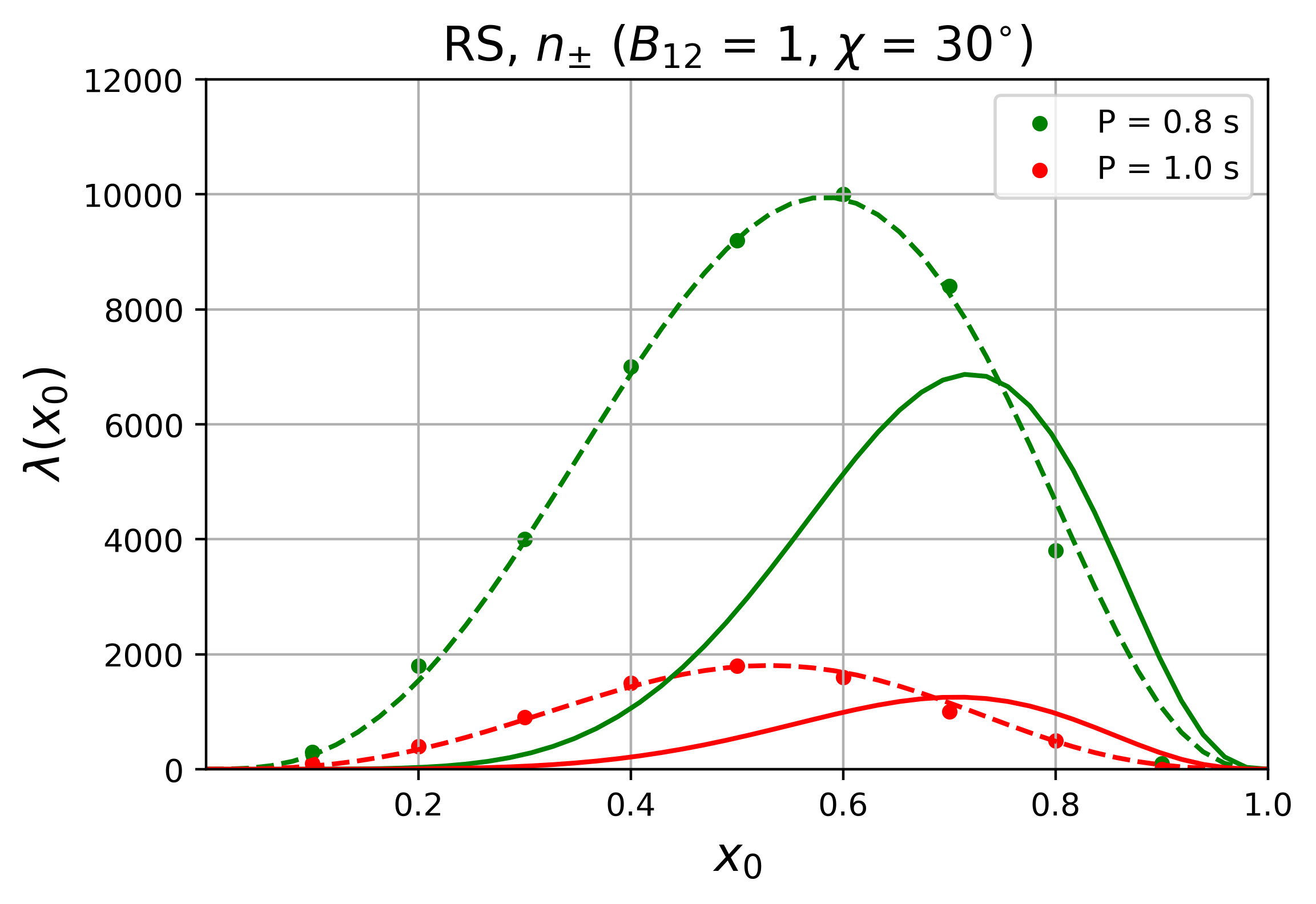}  }
	\caption{Total multiplication parameter $\lambda(x_{0}) = \lambda_{I} + \lambda_{II}$ 
             of secondary particles of the first and second generation   
             for pulsars with periods \mbox{$P = 0.8$ s} and $P = 1$ s as a function of 
             coordinate $x_{0}$ of a single primary particle. The dashed lines show how
             well the relation $\lambda \propto x_{0}^{3}$ holds for small $x_{0}$.
             Solid lines show how the parameter $<\omega_{\rm pe}^2/\gamma^3>$ 
             depends on $x_{0}$.}
\label{FigF}	
\end{figure}

Further, Table~\ref{tab3} shows the multiplication parameters $\lambda_{I}$ and $\lambda_{II}$ 
for two generations I and II for different values $x_{0}$ of a single primary particle.
We see that if for sufficiently fast pulsars ($P = 0.8$ s) the multiplication parameter 
$\lambda_{II}$ of the second generation significantly exceeds the multiplication parameter 
$\lambda_{I}$ of the first generation, for pulsars located near so-called ''death line'' 
($P = 1.2$ s), they become comparable to each other. In the latter case, the magnetosphere 
becomes transparent for most synchrotron photons, and therefore, despite their 
numerous, only a small part leads to the production of secondary pairs.

Besides, Figure~\ref{FigF} shows the total multiplication parameter $\lambda  = \lambda_{I} + \lambda_{II}$ 
for pulsars with periods \mbox{$P = 0.8$ s} and $P = 1$ s as a function of coordinate 
$x_{0}$ of a primary particle. As one can see, this sum models well enough the total 
number density $n_{\pm} = \lambda g(x_{0})n_{\rm GJ}$ of the secondary plasma. The dashed lines corresponding to 
fitting functions $f(x_{0}) \propto x_{0}^3 \exp(-x_{0}^2/x_{a}^2)$ show how 
well the relation $\lambda(x_{0}) \propto x_{0}^{3}$ holds for small  $x_{0}$. Accordingly, 
the solid lines show how the parameter $<\omega_{\rm pe}^2/\gamma^3>$ depends on $x_{0}$.

\begin{figure}
		\center{\includegraphics[width=0.9\linewidth]{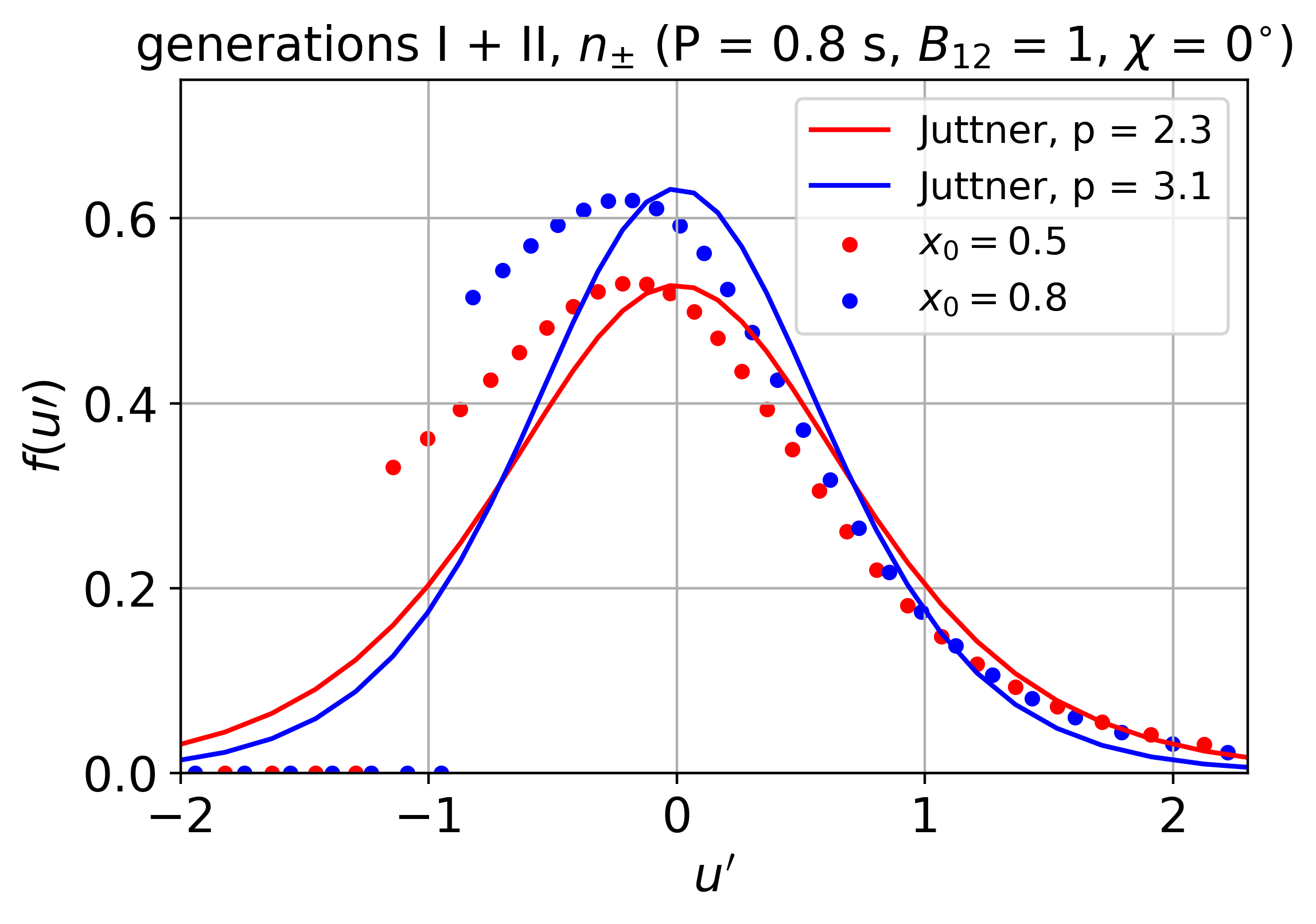}  }
	\caption{Distributions of the secondary particles as a function of their 
	longitudinal 4-velocity $u^{\prime}$ in the plasma rest frame for two 
	values $x_{0}$. Curve lines correspond to J{\" u}ttner distributions (\ref{Juttner}) with parameters $p$.}
\label{FigG}	
\end{figure}

Finally, Fig~\ref{FigG} shows the total (generations I and II) distributions 
of the secondary particles as a function of their longitudinal 4-velocity
$u^{\prime}$ in the plasma rest frame \mbox{($<v^{\prime}> \, = 0$)} for two 
values $x_{0} = 0.5$ and $0.8$.  Curve lines correspond to J{\" u}ttner distributions~\citep{Melrose1}
\begin{equation}
F(u^{\prime}) = \frac{e^{-p\gamma^{\prime}}}{2K_{1}(p)},
\label{Juttner}
\end{equation}
with appropriate parameters $p$. Here $\gamma^{\prime} = [1 + (u^{\prime})^{2}]^{1/2}$ and $\int F(u^{\prime}) {\rm d} u^{\prime} = 1$.  As we see, although the obtained distributions are close to the J{\" u}ttner distributions, they differ in noticeable asymmetry~\citep{M&B21}. Wherein, for all values of $x_{0}$ (and for pulsar periods $P \approx 1$ s under consideration), the parameters $p$ are within $1 < p < 5$, in full agreement with the results obtained by~\citet{AE2002}. This implies that the temperature of the secondary plasma $T$ is still less than the rest particle energy: $T < m_{\rm e}c^2$.

\section{Formation of triple profiles}
\label{Sect5}

Having determined, as we hope with sufficient accuracy, the number density of the outflowing 
plasma, we can proceed to our main task, namely, to determine the mean profiles formed by 
O-mode, taking into account its possible refraction due to a significant decrease in the 
density of the secondary plasma near the magnetic axis. Here we follow~\citet{HB14} who 
developed a method of determining the shape of the radio image in the picture plane as a 
function of pulse  phase $\phi$. It gives the possibility to study the changes of the size 
and the motion of the image along the picture plane. As to the shape of the mean profile, it 
can easily be obtained by integrating the intensity of the image in the picture plane. In 
addition, various locations of the emission region $r_{\rm rad}$ and energetic spectrum 
of the secondary plasma can also be considered.

Thus, following~\citealt{HB14}, we assume that the intensity of radio emission is proportional
to the plasma number density $n_{\rm e}(\bmath{r})$ at the radiation point $\bmath{r}$ which, in 
turn, can be easily determined knowing the number density profile (\ref{2}) determined above.
Indeed, the noticeable refraction of the O-mode takes place only at small distances from the 
star $r < r_{\rm O}$, where~\citep{B&A, BGI88}
\begin{equation}
r_{\rm O} \sim 10^2 R \, \lambda_{4}^{1/3} \gamma_{100}^{1/3} B_{12}^{1/3} \nu_{\rm GHz}^{-2/3} P^{1/5}.
\end{equation}
For the pulsars we are considering, these distances are hundreds of times less than the 
radius of the light cylinder $R_{\rm L} = c/\Omega \sim 10^4 R$. This allows us to restrict 
ourselves to the model of rigidly rotating dipole which, in turn, makes it trivial to find
the number density $n_{\rm e}$ at any point within open field line region. Finally, as for 
all O-mode pulsars listed in Table~\ref{tab0} the inclination angles are not too close to
$90^{\circ}$, we can assume the axisymmetric distribution of the number density of the 
outfowing plasma, i.e., that $g = g(r_{\perp})$. 

\begin{figure}
	\begin{minipage}{0.9\linewidth}
		\center{\includegraphics[width=0.8\linewidth]{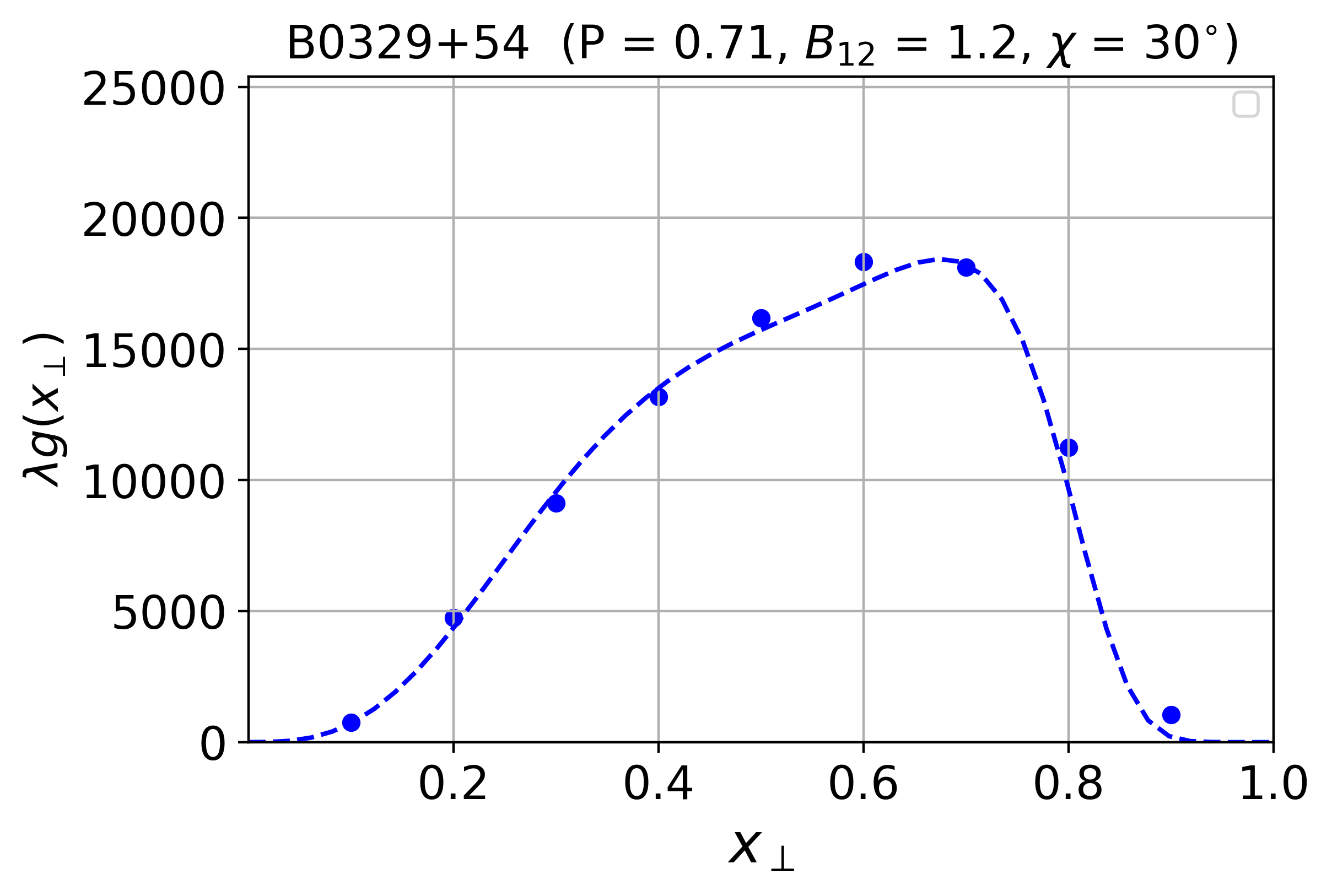} }
	\end{minipage}
	\hfill
	\begin{minipage}{0.9\linewidth}
		\center{\includegraphics[width=0.8\linewidth]{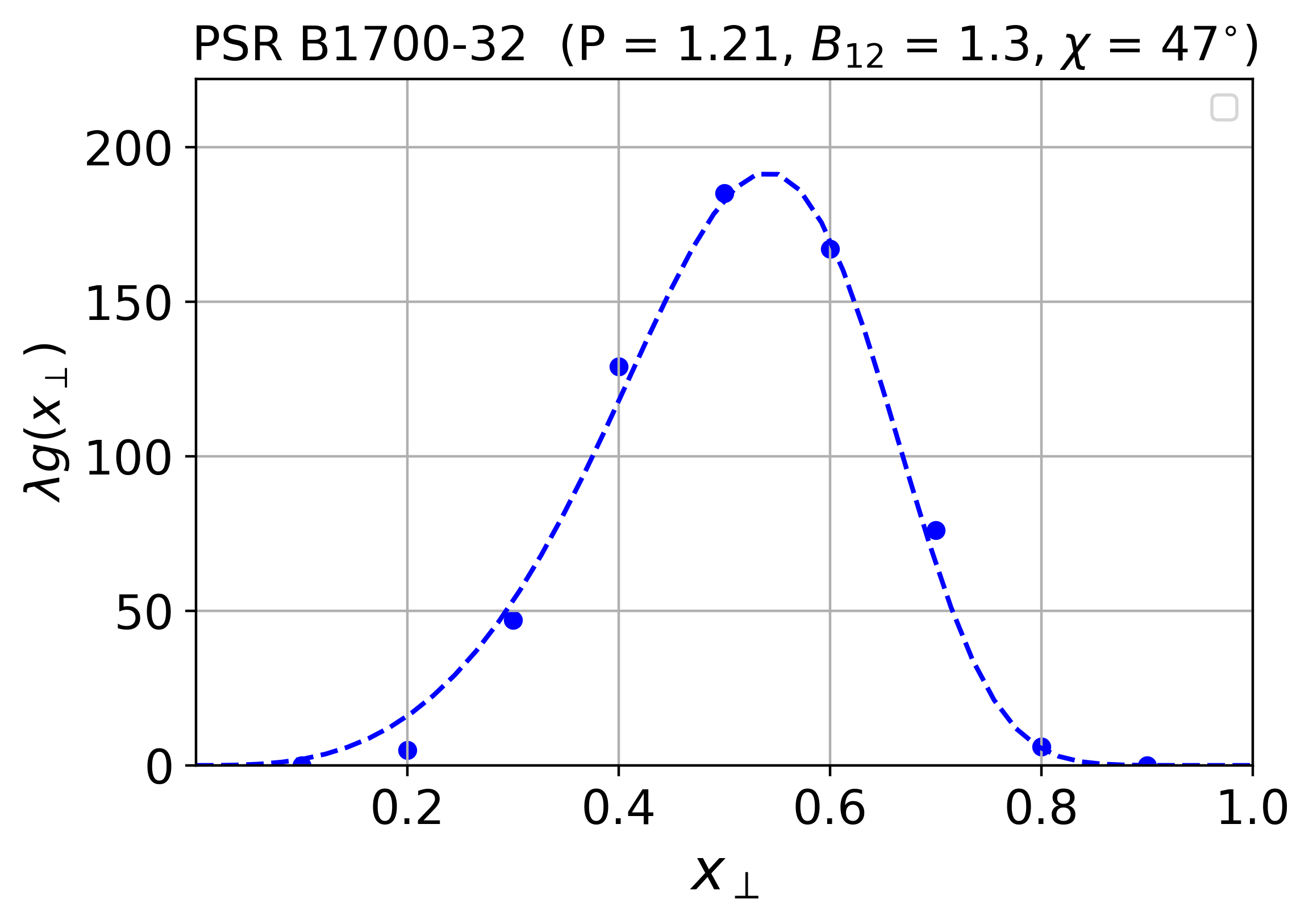}  }
	\end{minipage}
	\hfill
	\begin{minipage}{0.9\linewidth}
		\center{\includegraphics[width=0.8\linewidth]{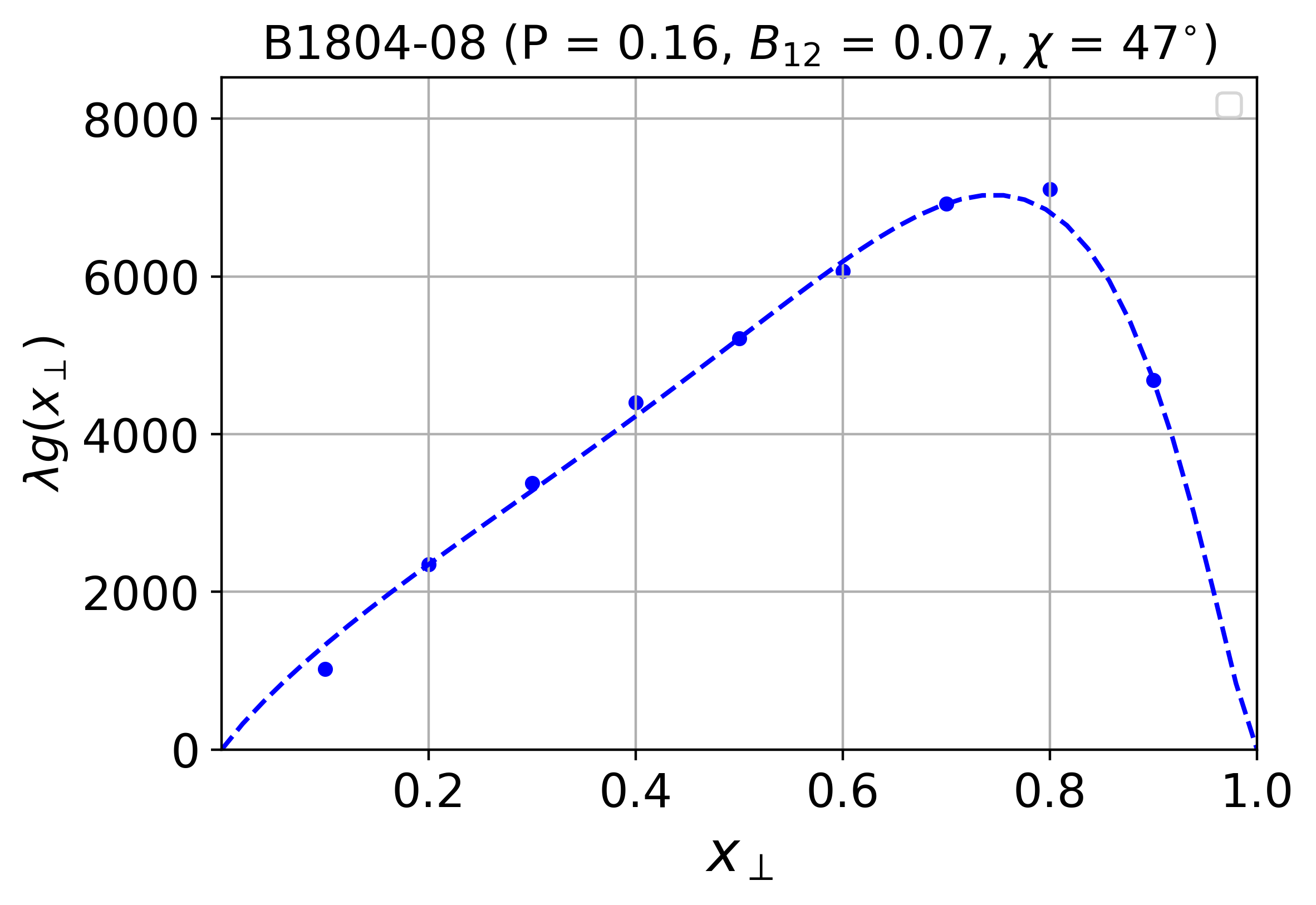} }
	\end{minipage}
	\vfill
	\begin{minipage}{0.9\linewidth}
		\center{\includegraphics[width=0.8\linewidth]{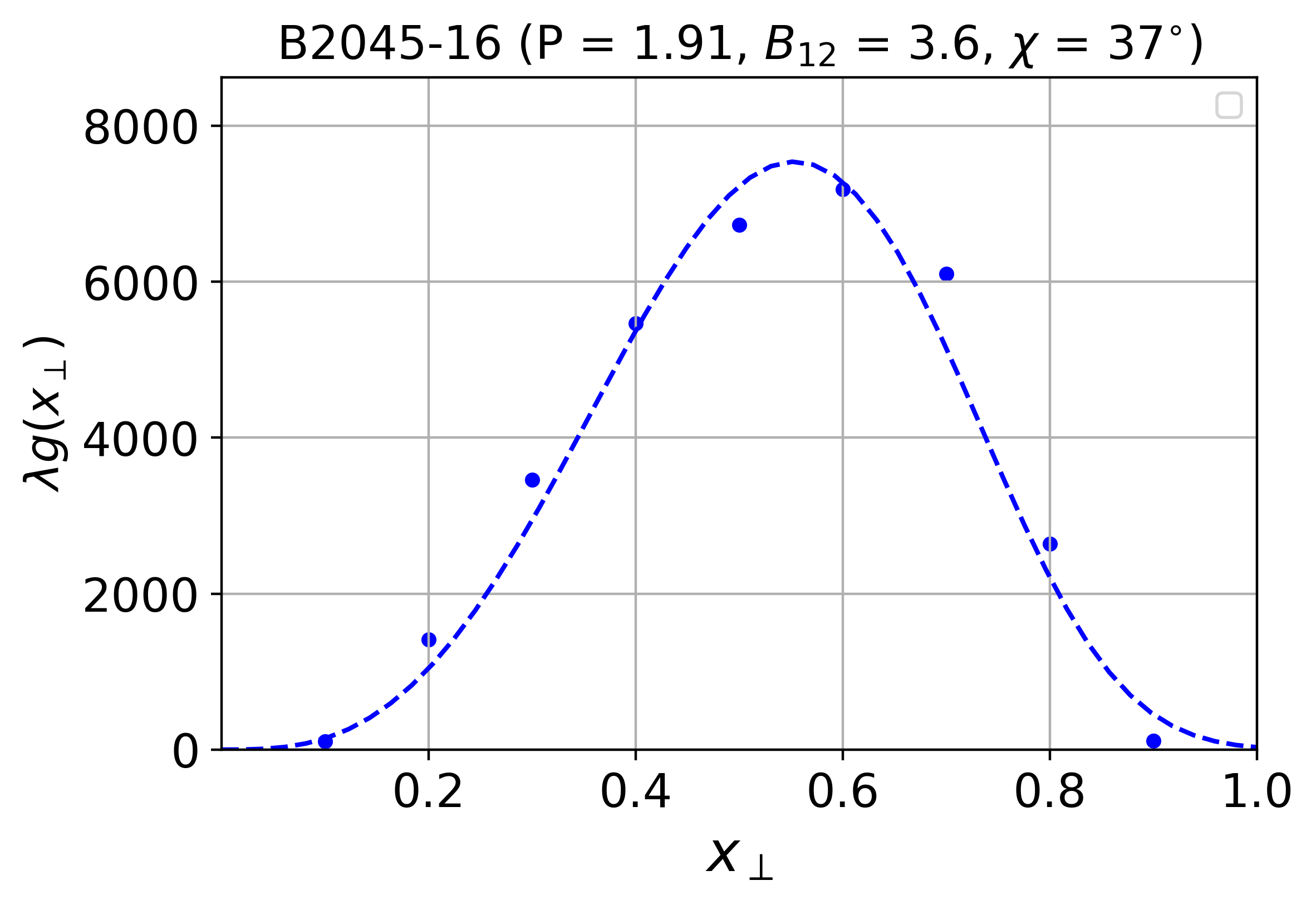}  }
	\end{minipage}
	\hfill
	\begin{minipage}{0.9\linewidth}
		\center{\includegraphics[width=0.8\linewidth]{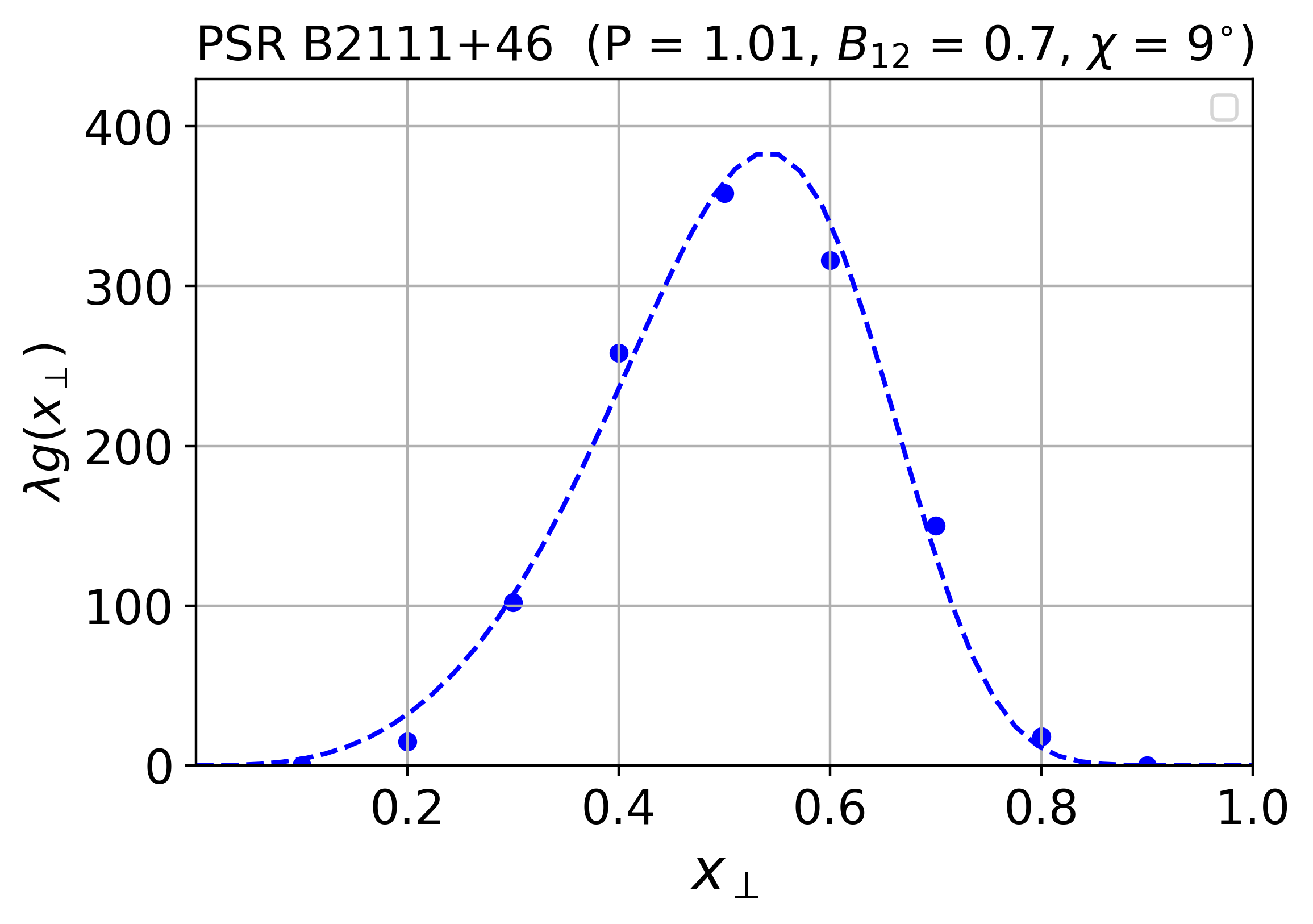}  }
	\end{minipage}
	\caption{Density profiles $\lambda g(r_{\perp}) = \lambda_{I} + \lambda_{II}$ of five O-mode pulsars listed 
	in Table~\ref{tab4}.}
\label{figH}	
\end{figure}

Accordingly, the refraction of the O-mode and, hence, the mean profile of the observed  
radiation also depend significantly on the plasma number density. Thus, the density profile  
of the outflowing plasma (\ref{2}) is our first most important parameter of the problem under
consideration. We emphasize once again that, in contrast to all previous works, the density 
profile discussed here is not a free parameter, but is determined self-consistently.

Figures~\ref{figH} show the density profiles $\lambda g(r_{\perp}) = \lambda_{I} + \lambda_{II}$ 
of five O-mode pulsars listed in Table~\ref{tab4}. Dashed lines correspond to fitting curves used 
in further calculations. As we see, in all cases the dependence $g(r_{\perp}) \propto r_{\perp}^3$ 
for small $r_{\perp}$ is fulfilled with good accuracy. As for the violation of this relation for 
two pulsars B0329$+$4544 and B1804$-$08, this property, as was already noted, is due to their small 
periods $P$. In addition, Table~\ref{tab4} contains the values of the multiplicities $\lambda$ and 
the factors $k$ (\ref{12}) which determine dependency $\gamma (r_{\perp}) \propto r_{\perp}^{- 1}$ 
(\ref{12}); this relation, as shown above, is executed with good accuracy. 

\begin{figure}
	\begin{minipage}{0.9\linewidth}
		\center{\includegraphics[width=0.8\linewidth]{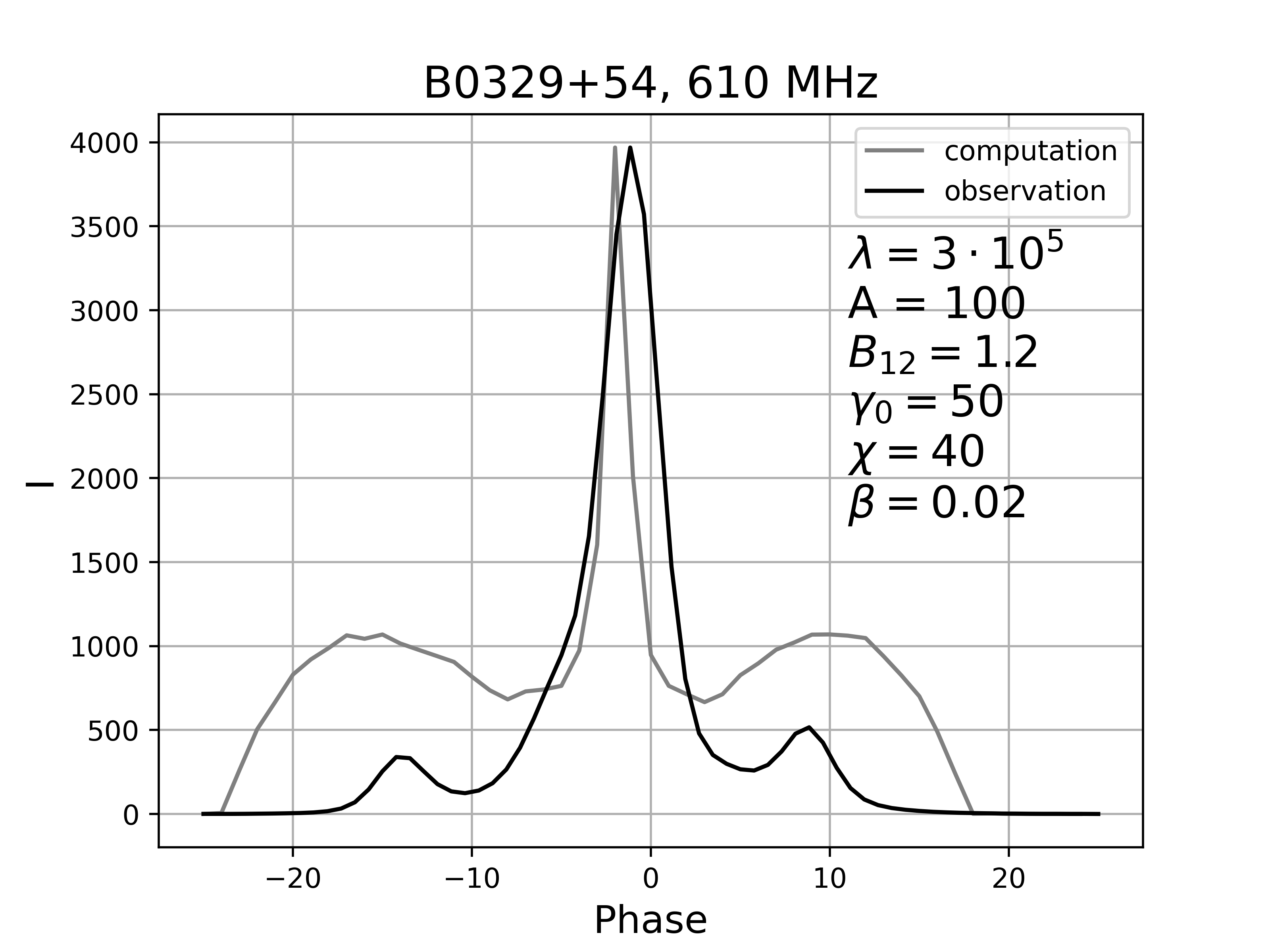} }
	\end{minipage}
	\hfill
	\begin{minipage}{0.9\linewidth}
		\center{\includegraphics[width=0.8\linewidth]{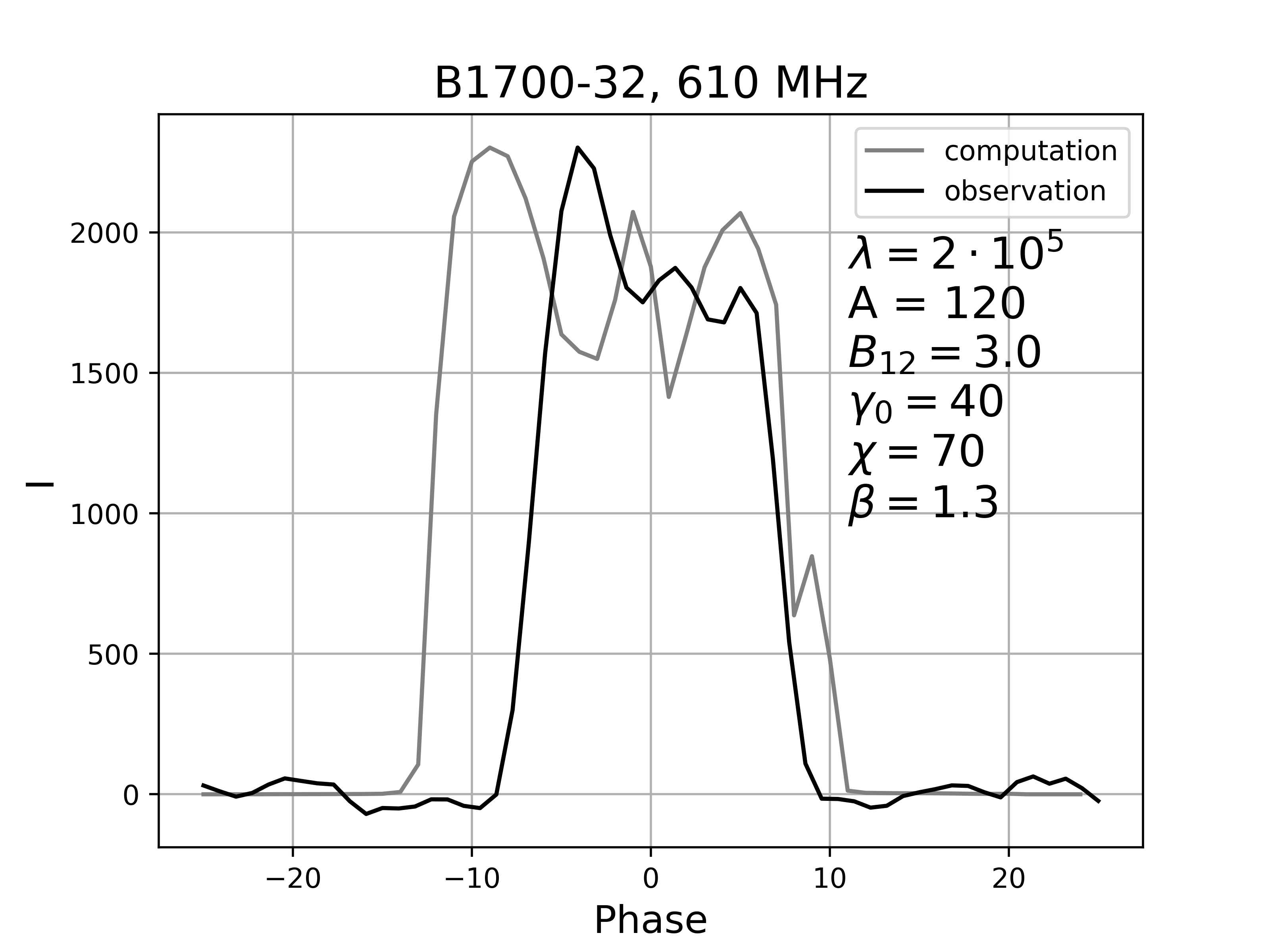}  }
	\end{minipage}
	\hfill
	\begin{minipage}{0.9\linewidth}
		\center{\includegraphics[width=0.8\linewidth]{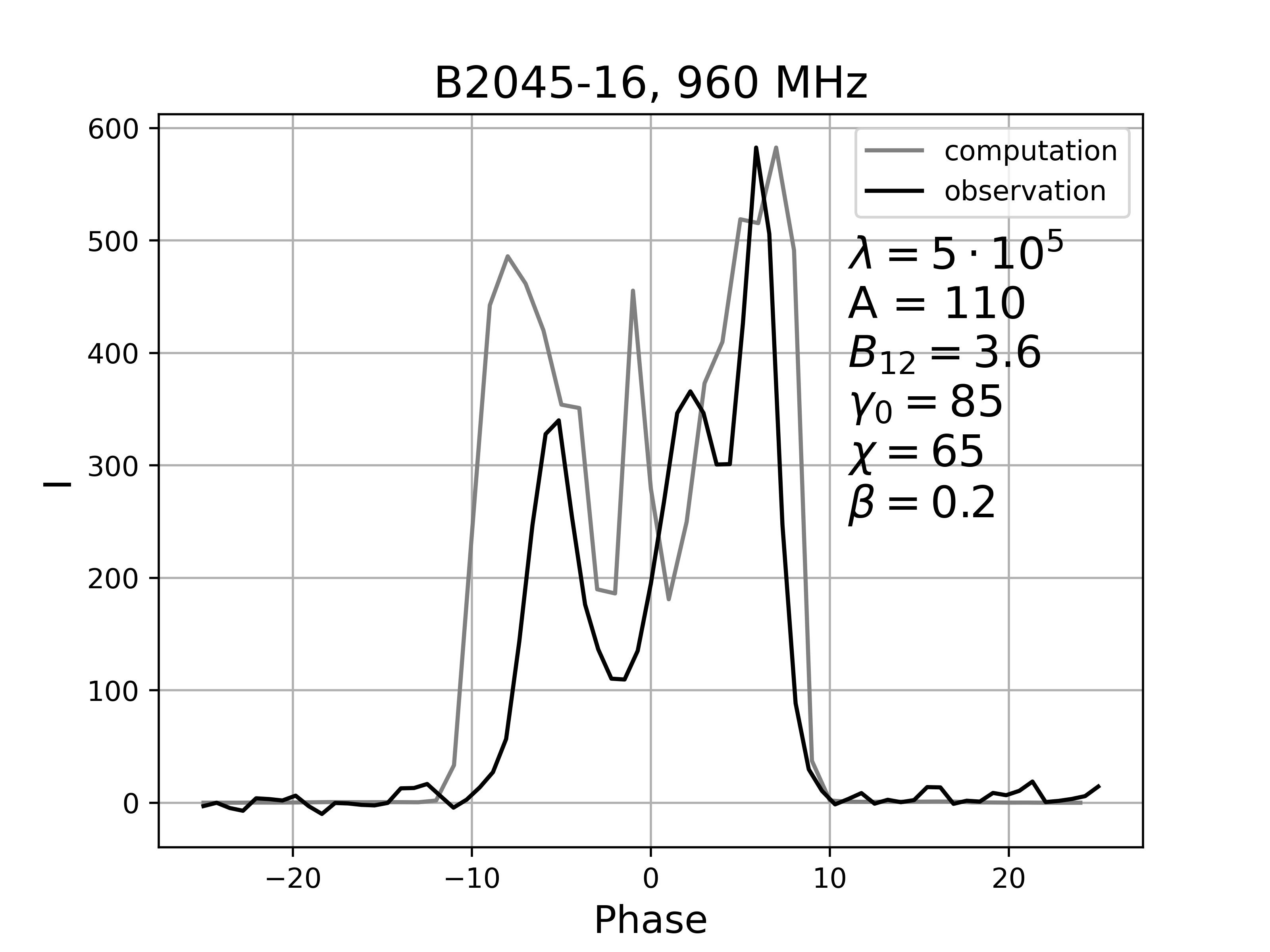} }
	\end{minipage}
	\vfill
	\begin{minipage}{0.9\linewidth}
		\center{\includegraphics[width=0.8\linewidth]{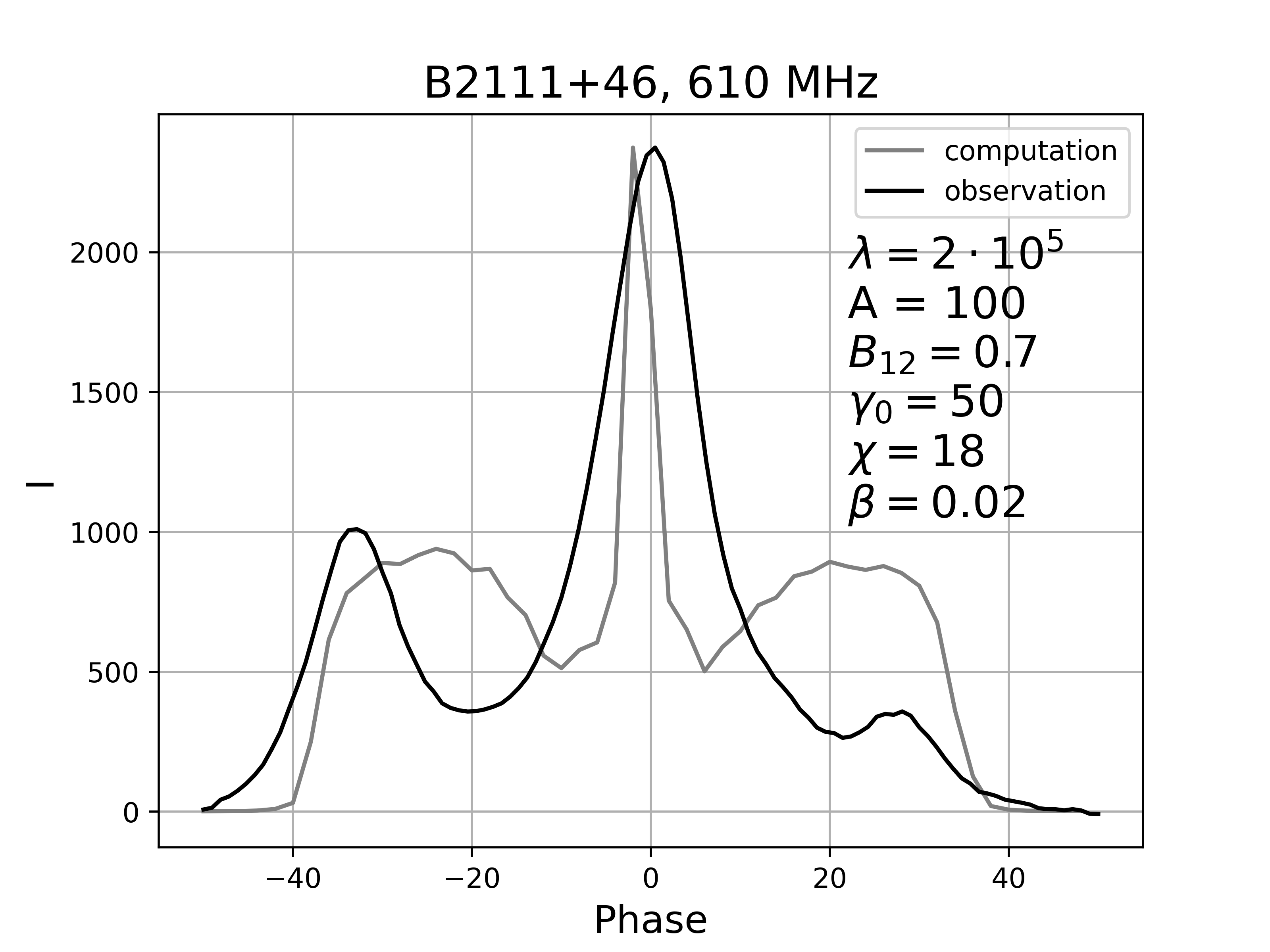}  }
	\end{minipage}
	\hfill
	\begin{minipage}{0.9\linewidth}
		\center{\includegraphics[width=0.8\linewidth]{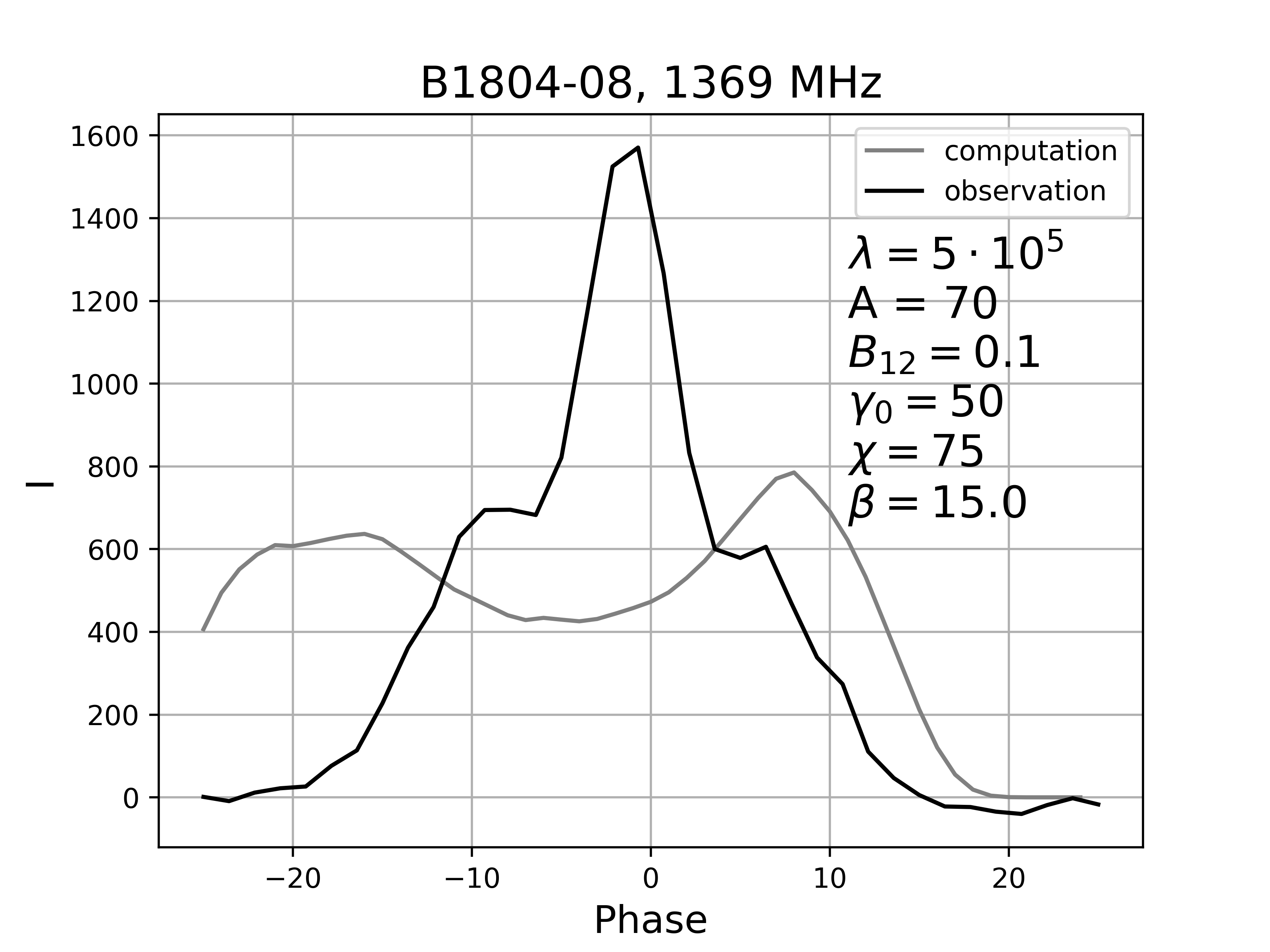}  }
	\end{minipage}
	\hfill
	\caption{Comparison of the computed profiles with the observed 
                  profiles of pulsars, whose three-hump profile is formed
                  by the ordinary O-mode. For each pulsar, all used 
                  parameters are indicated.}
\label{figI}	
\end{figure}

The second key parameter is the width of the beam for each radiating element relative 
to the direction of the magnetic field, $\theta_{\rm rad}$. We took this angle to be 
$\theta_{\rm rad} = \gamma^{-1}(r_{\perp})$, where $\gamma(r_{\perp})$ is the mean 
Lorentz-factor corresponding to given magnetic field line. Note that the self-consistent 
dependence of the Lorentz-factor on $r_{\perp}$ is also taken into account here for the 
first time. Thus, as we associate the observed radio emission with the outflowing secondary 
plasma, we assume in what follows that the radiation intensity is proportional to 
\begin{equation}
d(\theta_{\rm b}) = \exp(-\gamma^{2}\theta_{\rm b}^2),
\label{d_theta}
\end{equation}
where again $\theta_{\rm b}$ is the angle between the magnetic field line and the 
direction of the beam propagation. 

Thus, due to the factor $d(\theta)$ emission intensity of a chosen ray can be gained only if a ray almost touches a field line. So, if one assumes that all radio emission is produced on a fixed height, only a small fraction of open field line region will be able to emit towards observer. In this case mean intensity profiles should be heavily distorted and overall luminosity should be too weak. Therefore a broad emission region model was chosen and an additional cutoff factor which models inability to produce radiation far form the star was introduced in a form:
\begin{equation}
h(r) = \exp \left[-\frac{r^2}{A^2 R^2}\right].
\end{equation}
Here $A \gg 1$ which implies a fairly extensive emission region. It should be stressed once again that while there is no specific emission height in this model, each ray gains its intensity in a small region around specific height determined the by the field lines geometry due to $d(\theta)$ factor.

If one also assumes that emission intensity in a given point of magnetosphere is proportional to a plasma density, overall intensity from a one ray can be calculated as follows:
\begin{equation}
I =\int_{\mathbf{r}(l)} g\left(r_{\perp}\right) h(r) d(\theta) d l
\label{intensity}
\end{equation}
Here an integral is taken on a ray trajectory.

Considering now parallel rays that simultaneously intersect a screen (the picture 
plane) perpendicular to the direction toward the observer, we can, by integrating
equations of geometric optics 
\begin{eqnarray}
\frac{{\rm d}\bmath{r}}{{\rm d}l} & = & \frac{\partial (k/n_{2})}{\partial \bmath{k}}, \nonumber \\
\frac{{\rm d}\bmath{k}}{{\rm d}l} & = & -\frac{\partial (k/n_{2})}{\partial \bmath{r}},
\end{eqnarray}
back to the rotating neutron star, determine the relative intensity of the rays 
passing through various points of the screen. Here, the refractive index 
$n_{2} = n_{2}(n_{\rm e}, \theta_{\rm b})$ is determined by the relation (\ref{1}). Wherein, 
different times of the beginning of integration can be easily associated with the 
observed phase of the pulse $\phi$. In this work, we do not discuss the shape of the
image itself and its motion in the picture plane (see~\citealt{HB14} for more detail), 
giving only the dependence of the total intensity on the phase of the pulse $\phi$.
It should also be mentioned that here we do not take into account cyclotron absorption.

\begin{table*}
\caption{Parameters of triple O-mode pulsars discussed in this paper.}
\begin{tabular}{cccccc}
\hline
    & B0329$+$54 &   B1700$-$32   &  B1804$-$08 &  B2045$-$16  &  B2111$+$46    \\
\hline
$P$(s)           & 0.71 & 1.21 & 0.16 & 1,96 & 1.01 \\
${\dot P}_{-15}$ & 2.05 & 0.66 & 0.03 & 11.0 & 0.71 \\
$\chi(^\circ)$   &  30  &  47  &  47  &  46 &  9 \\
$B_{12}$         &  1.2 &  1.3 & 0.07 & 3.6 & 0.7 \\ 
$\lambda$        & $10^4$ & 60 & 4000 & 3500 & 110 \\
$ k (\ref{12})$  &  1.3 &  1.1 &  1.2  & 1.1 & 1.2 \\
\hline
\end{tabular}
\label{tab4}
\end{table*}

Using all the above results now, we are ready to determine the mean profile for each of the five pulsars. They are shown in Figure~\ref{figI}. As one can see, in four out of five cases there is excellent agreement between the synthesized and observed three-hump profiles. And this is despite the fact that we have considered a fairly simple model of the generation of the radio emission.
 Hence, we can confidently conclude that the central peak in triple pulsars, whose mean profile is formed by the ordinary O-mode, can easily be explained by its refraction in an inhomogeneous plasma outflowing along open magnetic field lines in the pulsar magnetosphere.

As for the parameters used in modelling, we should make some clarifications. First of all, multiplicity parameters were taken sufficiently larger than it follows from the theoretical calculation. That is due to the fact that only two generations of secondary particles were taken into account. From the numerical calculations it follows that multiplicity should be $\sim 10^5$ to produce significant mean hump. The other parameters used in program only slightly differ from given in Table~\ref{tab4}. 

Another uncertainty was related to inclination angle $\chi$. As its value is almost impossible to determine with enough precision it was chosen to match profile width with observational data. Finally, a magnetic field is also depend on $\chi$ angle and thus can be only estimated. 

\section{Conclusion}

Thus, in the development of the idea proposed by~\citet{P&L1, P&L2}, we have shown that the triple mean profiles of radio pulsars can be easily explained by the refraction of the ordinary O-mode in the pulsar magnetosphere. The essential advance here was that we used much more realistic particle number density and energy profiles than have been done so far.

It should be stated once again, that the main goal of this numerical study was to show qualitative similarity between observations and proposed theory. Quantitative correspondence is not feasible within this framework because some important effects were not considered. First of all, it does not follow from anywhere that the radio emission profile repeats the density of the emitting plasma. Further, the simplest model of a rotating dipole was used, which is obviously not valid for millisecond pulsars. Finally, it must be mentioned that here we did not take into account cyclotron absorption, Which, as is well known~\citep{ML04, beskinphilippov2012}, can significantly distort the mean profile. We intend to take into account all these circumstances in the next work.

Separately, it should be noted that a quadrupole~\citep{quad} or small-scale~\citep{Kantor, Barsukov} magnetic field can significantly affect the plasma density profile. However, this issue requires a separate detailed study. Therefore, here we confine ourselves to only one remark about the conditions for a significant change in the curvature of the magnetic field lines in the plasma generation region.

As shown in Appendix~\ref{A}, the influence of a quadrupole magnetic field remains negligible for small enough ratio $b_{q} = B_{q}/B_{0} < 0.1$. In this case, the field line passing through the 'magnetic pole' on the surface of the neutron star, at large distances, where the field can already be considered purely dipole, shifts only within 10\% of the size of the open field line boundary. On the other hand, for larger values of $b_{q} > 0.1$, the curvature radius $R_{\rm c}$ at the base of open field lines  becomes almost constant; in this case our consideration already ceases to be fair.

Summing up, we note that despite all the simplifications made in our work, we managed to achieve a fairly good agreement between the predictions of the theory and the observational data. We hope that in the future this line of research will indeed make it possible to reproduce the average profiles of radio pulsars with sufficient accuracy, and, consequently, to determine the physical conditions in the region of their generation.

\section*{Data availability}
The list of triple pulsars was taken from~\citet{RankinIV}.
Other data underlying this work will be shared on reasonable request to the corresponding author.

\section*{Acknowledgements}
Authors thank Hayk Hakobyan, Fedor Kniazev, Sasha Philippov and Denis Sob'yanin for useful discussions. 
This work was partially supported by the National Research Center 'Kurchatov Institute' (Order No. 85 dated 03.20.23).

\bibliographystyle{mnras}
\bibliography{Omode}

\appendix

\section{Quadrupole correction}
\label{A}

In this Appendix we consider only the plane containing the dipole and quadrupole axes; as the magnetic field lines also lie in this plane, the question of their curvature is greatly simplified. Indeed, for a plane curve, the curvature radius of the magnetic field line can be written as~\citep{Korns}
\begin{eqnarray}
R_{\rm c} = 1/|(\bmath{h}\nabla)\bmath{h}| =  1/|\bmath{h}\times [\nabla \times \bmath{h}]| 
\nonumber \\
= \frac{r}{|h_{\theta} + r\partial h_{\theta}/\partial r - \partial h_{r}/\partial \theta|},
\label{A1}
\end{eqnarray}
where $\bmath{h}$ is the unit vector along the magnetic field line \mbox{($\bmath{h} = \bmath{B}/B$).} Using now the explicit expressions for the dipole and quadrupole magnetic fields
\begin{eqnarray}
&&B_{r} = B_{0}\cos\theta \frac{R^3}{r^3} + B_{q}[3\cos^2(\theta - \theta_{0}) -1]\frac{R^4}{2r^4},
\label{A2} \\
&&B_{\theta} = B_{0}\sin\theta \frac{R^3}{2r^3} + B_{q}\sin(\theta - \theta_{0})\cos(\theta - \theta_{0})\frac{R^4}{r^4},
\nonumber 
\end{eqnarray}
where $\theta_{0}$ is the inclination angle of the quadrupole axis to the dipole magnetic axis, it is possible to determine both the unit vector $\bmath{h}$ (and, hence, the curvature of the magnetic field lines) and the shape of the polar cap on the surface of the pulsar. Here we assume that the region of open field lines at large distances from the neutron star coincides with one for a pure dipole.

\begin{figure}
		\center{\includegraphics[width=0.9\linewidth]{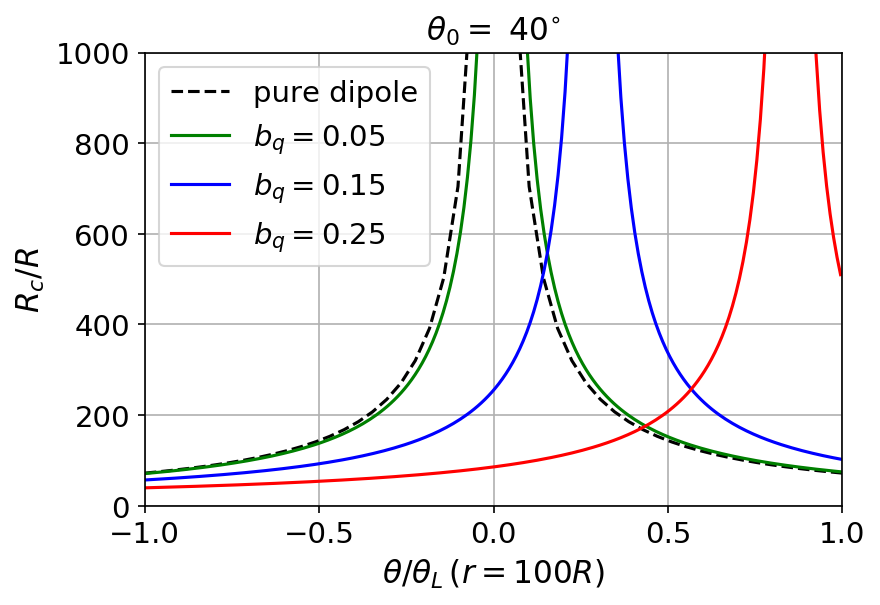}  }
	\caption{Curvature radius $R_{\rm c}$ at the base of the field lines (i.e., for $r = R$) as a function of their angle $\theta$ at large distance \mbox{$r = 100 \, R$,} where the field can be considered purely dipole. 
    }
\label{figApp}	
\end{figure}

As expected, the additional quadrupole magnetic field results in a shift of the 'zero' magnetic field line, i.e. that field line which has infinite curvature radius at its base $r = R$. As shown in Fig.~\ref{figApp}, significant shift occurs when the ratio $b_{\rm q} = B_{0}/B_{\rm q}$ exceeds 10\%. 

\end{document}